\documentclass[twocolumn]{aastex63}

\submitjournal{AAS Journals}
\usepackage{amsmath,amstext}
\usepackage[T1]{fontenc}
\usepackage{multirow}
\usepackage[utf8]{inputenc} 
\usepackage{float}
\usepackage{graphicx}
\newcommand{\unit}[1]{\ensuremath{\, \mathrm{#1}}}
\shorttitle{Two Planets Around the M-dwarf TOI-1266}
\shortauthors{Stefansson et al.}

\begin{document}

\title{A Mini-Neptune and a Venus-Zone Planet in the Radius Valley Orbiting the Nearby M2-dwarf TOI-1266: Validation with the Habitable-zone Planet Finder}
\correspondingauthor{Gudmundur Stefansson}
\email{gstefansson@astro.princeton.edu}

\author[0000-0001-7409-5688]{Gudmundur Stefansson}
\altaffiliation{Henry Norris Russell Fellow}
\affiliation{Princeton University, 4 Ivy Lane, Princeton, NJ 08540, USA}

\author[0000-0002-5893-2471]{Ravi Kopparapu}
\affiliation{NASA Goddard Space Flight Center, 8800 Greenbelt Road, Greenbelt, MD 20771, USA}

\author[0000-0002-9082-6337]{Andrea Lin}
\affil{Department of Astronomy \& Astrophysics, The Pennsylvania State University, 525 Davey Laboratory, University Park, PA, 16802, USA}
\affil{Center for Exoplanets and Habitable Worlds, 525 Davey Laboratory, University Park, PA, 16802, USA}

\author[0000-0001-9596-7983]{Suvrath Mahadevan}
\affil{Department of Astronomy \& Astrophysics, The Pennsylvania State University, 525 Davey Laboratory, University Park, PA, 16802, USA}
\affil{Center for Exoplanets and Habitable Worlds, 525 Davey Laboratory, University Park, PA, 16802, USA}

\author[0000-0003-4835-0619]{Caleb I. Ca\~nas}
\altaffiliation{NASA Earth and Space Science Fellow}
\affil{Department of Astronomy \& Astrophysics, The Pennsylvania State University, 525 Davey Laboratory, University Park, PA, 16802, USA}
\affil{Center for Exoplanets and Habitable Worlds, 525 Davey Laboratory, University Park, PA, 16802, USA}

\author[0000-0001-8401-4300]{Shubham Kanodia}
\affil{Department of Astronomy \& Astrophysics, The Pennsylvania State University, 525 Davey Laboratory, University Park, PA, 16802, USA}
\affil{Center for Exoplanets and Habitable Worlds, 525 Davey Laboratory, University Park, PA, 16802, USA}

\author[0000-0001-6160-5888]{Joe P.\ Ninan}
\affil{Department of Astronomy \& Astrophysics, The Pennsylvania State University, 525 Davey Laboratory, University Park, PA, 16802, USA}
\affil{Center for Exoplanets and Habitable Worlds, 525 Davey Laboratory, University Park, PA, 16802, USA}

\author[0000-0001-9662-3496]{William D. Cochran}
\affil{McDonald Observatory and Department of Astronomy, The University of Texas at Austin, 2515 Speedway, Austin, TX 78712, USA}
\affil{Center for Planetary Systems Habitability, The University of Texas at Austin, 2515 Speedway, Austin, TX 78712, USA}

\author[0000-0002-7714-6310]{Michael Endl}
\affil{McDonald Observatory and Department of Astronomy, The University of Texas at Austin}
\affil{Center for Planetary Systems Habitability, The University of Texas at Austin}

\author[0000-0003-1263-8637]{Leslie Hebb}
\affiliation{Department of Physics, Hobart and William Smith Colleges, 300 Pulteney Street, Geneva, NY, 14456, USA}

\author[0000-0001-9209-1808]{John Wisniewski}
\affiliation{Homer L. Dodge Department of Physics and Astronomy, University of Oklahoma, 440 W. Brooks Street, Norman, OK 73019, USA}

\author[0000-0002-5463-9980]{Arvind Gupta}
\affil{Department of Astronomy \& Astrophysics, The Pennsylvania State University, 525 Davey Laboratory, University Park, PA, 16802, USA}
\affil{Center for Exoplanets and Habitable Worlds, 525 Davey Laboratory, University Park, PA, 16802, USA}

\author[0000-0002-0885-7215]{Mark Everett}
\affiliation{NSF's OIR Lab, 950 N. Cherry Ave. Tucson, AZ 85719, USA}

\author[0000-0003-4384-7220]{Chad F.\ Bender}
\affil{Steward Observatory, The University of Arizona, 933 N.\ Cherry Ave, Tucson, AZ 85721, USA}

\author[0000-0002-2144-0764]{Scott A. Diddams}
\affil{Time and Frequency Division, National Institute of Standards and Technology, 325 Broadway, Boulder, CO 80305, USA}
\affil{Department of Physics, University of Colorado, 2000 Colorado Avenue, Boulder, CO 80309, USA}

\author[0000-0001-6545-639X]{Eric B.\ Ford}
\affil{Department of Astronomy \& Astrophysics, The Pennsylvania State University, 525 Davey Laboratory, University Park, PA, 16802, USA}
\affil{Center for Exoplanets and Habitable Worlds, 525 Davey Laboratory, University Park, PA, 16802, USA}
\affil{Institute for Computational \& Data Sciences, University Park, PA, 16802, USA}

\author[0000-0002-0560-1433]{Connor Fredrick}
\affil{Time and Frequency Division, National Institute of Standards and Technology, 325 Broadway, Boulder, CO 80305, USA}
\affil{Department of Physics, University of Colorado, 2000 Colorado Avenue, Boulder, CO 80309, USA}

\author[0000-0003-1312-9391]{Samuel Halverson}
\altaffiliation{Sagan Fellow}
\affil{MIT Kavli Institute for Astrophysics, 70 Vassar St, Cambridge, MA 02109, USA}
\affil{Jet Propulsion Laboratory, 4800 Oak Grove Drive, Pasadena, CA 91109, USA}

\author[0000-0002-1664-3102]{Fred Hearty}
\affil{Department of Astronomy \& Astrophysics, The Pennsylvania State University, 525 Davey Laboratory, University Park, PA, 16802, USA}
\affil{Center for Exoplanets and Habitable Worlds, 525 Davey Laboratory, University Park, PA, 16802, USA}

\author{Eric Levi}
\affil{Department of Astronomy \& Astrophysics, The Pennsylvania State University, 525 Davey Laboratory, University Park, PA, 16802, USA}
\affil{Center for Exoplanets and Habitable Worlds, 525 Davey Laboratory, University Park, PA, 16802, USA}

\author[0000-0001-8222-9586]{Marissa Maney}
\affil{Department of Astronomy \& Astrophysics, The Pennsylvania State University, 525 Davey Laboratory, University Park, PA, 16802, USA}

\author[0000-0001-5000-1018]{Andrew J. Metcalf}
\affiliation{Space Vehicles Directorate, Air Force Research Laboratory, 3550 Aberdeen Ave. SE, Kirtland AFB, NM 87117, USA}
\affiliation{Time and Frequency Division, National Institute of Standards and Technology, 325 Broadway, Boulder, CO 80305, USA} 
\affiliation{Department of Physics, University of Colorado, 2000 Colorado Avenue, Boulder, CO 80309, USA}

\author[0000-0002-0048-2586]{Andrew Monson}
\affil{Department of Astronomy \& Astrophysics, The Pennsylvania State University, 525 Davey Laboratory, University Park, PA, 16802, USA}
\affil{Center for Exoplanets and Habitable Worlds, 525 Davey Laboratory, University Park, PA, 16802, USA}

\author[0000-0002-4289-7958]{Lawrence W. Ramsey}
\affil{Department of Astronomy \& Astrophysics, The Pennsylvania State University, 525 Davey Laboratory, University Park, PA, 16802, USA}
\affil{Center for Exoplanets and Habitable Worlds, 525 Davey Laboratory, University Park, PA, 16802, USA}

\author[0000-0003-0149-9678]{Paul Robertson}
\affil{Department of Physics and Astronomy, The University of California, Irvine, Irvine, CA 92697, USA}

\author[0000-0001-8127-5775]{Arpita Roy}
\altaffiliation{Robert A. Millikan Postdoctoral Fellow}
\affil{Department of Astrophysics, California Institute of Technology, Pasadena, CA 91125, USA}

\author[0000-0002-4046-987X]{Christian Schwab}
\affil{Department of Physics and Astronomy, Macquarie University, Balaclava Road, North Ryde, NSW 2109, Australia}

\author[0000-0002-4788-8858]{Ryan C. Terrien}
\affil{Department of Physics and Astronomy, Carleton College, One North College Street, Northfield, MN 55057, USA}

\author[0000-0001-6160-5888]{Jason T.\ Wright}
\affil{Department of Astronomy \& Astrophysics, The Pennsylvania State University, 525 Davey Laboratory, University Park, PA, 16802, USA}
\affil{Center for Exoplanets and Habitable Worlds, 525 Davey Laboratory, University Park, PA, 16802, USA}

\begin{abstract}
We report on the validation of two planets orbiting the nearby ($36 \unit{pc}$) M2 dwarf TOI-1266 observed by the TESS mission. The inner planet is sub-Neptune-sized ($R=2.46 \pm 0.08 R_\oplus$) with an orbital period of 10.9 days. The outer planet has a radius of $1.67_{-0.11}^{+0.09} R_\oplus$ and resides in the exoplanet Radius Valley---the transition region between rocky and gaseous planets. With an orbital period of 18.8 days, the outer planet receives an insolation flux of 2.4 times that of Earth, similar to the insolation of Venus. Using precision near-infrared radial velocities with the Habitable-zone Planet Finder Spectrograph, we place upper mass limits of $15.9 M_\oplus$ and $6.4 M_\oplus$ at 95\% confidence for the inner and outer planet, respectively. A more precise mass constraint of planet c, achievable with current RV instruments given the host star brightness (V=12.9, J=9.7), will yield further insights into the dominant processes sculpting the exoplanet Radius Valley.
\end{abstract}

\section{Introduction}
\label{sec:intro}
One of the key findings from the \textit{Kepler} mission \citep{borucki2010} is that planets with radii between Earth ($1 R_\oplus$) and Neptune ($4 R_\oplus$)---which are not known to exist in the Solar System---are prevalent \cite[e.g.,][]{howard2012,fressin2013,batalha2013,petigura2013,dressing2015}. In this grouping of planets, \textit{Kepler} data further showed convincing evidence that there is a dip in the radius distribution of Kepler planets at 1.5-2.0 Earth radii \citep{owen2013,fulton2017,vaneylen2018}. This gap, or `Radius Valley', has been interpreted as the transition between predominantly rocky planets (super-Earths) populating the space below the gap, and planets rich in volatiles or ices residing above the gap (sub-Neptunes). Subsequent studies have found evidence of the Radius Valley in the \textit{K2} mission \citep{hardegree2020}, and have also explored how it varies as a function of stellar type \cite[e.g.,][]{cloutier2020}.

The astrophysical origin of the Radius Valley has been explored by a number of groups \citep[see e.g.,][]{owen2013,lee2014,owen2017,lopez2018}. Different theoretical models predict that the location of the rocky-to-gaseous transition radius should depend on the planet orbital period. Among these, photoevaporation \citep{lopez2012,owen2013,lopez2013,owen2017}---where a planet's primordial atmosphere is stripped by XUV photons from the host star---predicts that the rocky-to-gaseous transition radius should decrease with orbital period (as $\sim$$P^{-0.15}$). Second, internally-driven thermal atmospheric escape models via the core-powered mass loss mechanism \citep{ginzburg2016,ginzburg2018,gupta2019} also predict that the location of the Radius Valley should decrease with orbital period (as $\sim$$P^{-0.13}$). Third, giant impacts can also provide a way to sculpt the atmospheric properties of small planets and strip large primordial envelopes down to a few percent by mass \citep{inamdar2015,liu2015}. Conversely, models assuming formation at later times in a gas-poor environment \citep{lee2014,lee2016,lopez2018} predict that the location of the Radius Valley should increase with period (as $\sim$$P^{0.11}$).

Knowledge of planetary bulk densities---and thus planetary compositions---as a function of orbital period, offers a direct observational test of the predictions of the different hypotheses mentioned above. However, the current number of planets with precise bulk density constraints are insufficient to robustly identify the dominant formation pathway of the Radius Valley \citep{cloutier2020}. The Transiting Exoplanet Survey Satellite \citep[TESS;][]{Ricker2014}, which is surveying the night sky for transiting exoplanets around the nearest and brightest stars, is finding more planets amenable to precise mass measurements. 

We report on the discovery and ground-based validation of two small exoplanets orbiting the nearby M-dwarf TOI-1266 observed in four Sectors of TESS data. The inner planet has a period of $P=10.9$ days and radius of $R=2.46 \pm 0.08 R_\oplus$, and likely has a gaseous envelope. The outer planet has a period of $P=18.8$ days and radius of $R=1.67_{-0.11}^{+0.09}  R_\oplus$, and thus resides in the exoplanet Radius Valley, and could either have retained a small gaseous envelope or have a predominantly rocky composition. Receiving insolation fluxes of $4.7_{-0.7}^{+1.0}S_\oplus$, and $2.42_{-0.22}^{+0.23}S_\oplus$, both planets reside in the exoplanet 'Venus-Zone'---the region between the runaway greenhouse boundary defined by \cite{kopparapu2013} and $25 S_\oplus$ \citep{kane2014,ostberg2019}---where the outer planet has an insolation flux similar to that of Venus of $1.91S_\oplus$. The detailed characterization of systems in the Venus-Zone, including mass and atmospheric compositions, will increase our understanding of the limits of habitable environments. Using precise radial velocities from the Habitable-zone Planet Finder Spectrograph, we place upper limits on the mass of both planets. Both planets are amenable for mass constraints with additional RV observations. A mass constraint of the outer planet will allow its composition to be determined, and will be a valuable data point in discerning between competing models explaining the emergence of the Radius Valley.

This paper is structured as follows. Section \ref{sec:obs} describes the observations and data reduction. In Section \ref{sec:stellarparams}, we describe the key parameters of the host star, and in Section \ref{sec:planetparams} we describe our constraints on parameters of the planets. In Section \ref{sec:validation}, we statistically validate both planets. In Section \ref{sec:discussion}, we place the TOI-1266 system in context with other exoplanet systems, and we conclude in Section \ref{sec:summary} with a summary of our key findings.

\section{Observations and Data Reduction}
\label{sec:obs}

\begin{figure*}[t!]
\begin{center}
\includegraphics[width=\textwidth]{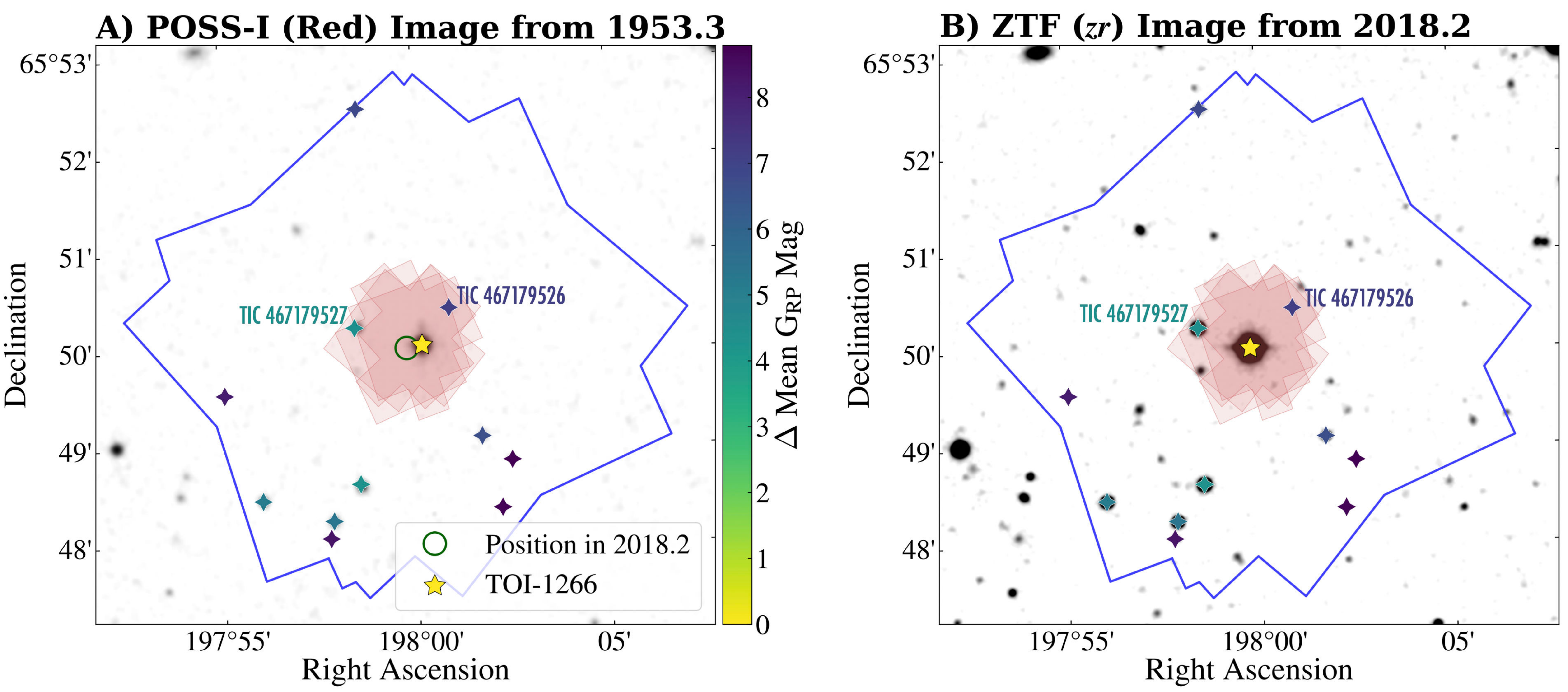}
\end{center}
\caption{TESS apertures (red shaded areas) and full TESS $11 \times 11$ pixel grids (blue lines) highlighted over seeing-limited images from a) POSS-1 from 1955.3, and b) the Zwicky Transiet Facility \citep[ZTF;][]{masci2019} from 2018.2. The location of TOI-1266 is noted with the yellow star. Other nearby stars as detected by Gaia are highlighted with the colorbar. Two nearby stars that are partially overlapping with the TESS aperture are highlighted, but due to their faintness, they result in minimal dilution of the TESS light curve. The green circle in a) highlights the position of TOI-1266 in 2018, showing no evidence of an overlapping background star at its position during the TESS observations.}
\label{fig:fov}
\end{figure*}

\subsection{TESS Photometry}
TOI-1266 was observed by TESS in 4 sectors in Sector 14 (Camera 4; July 18, 2019 -- August 15, 2019), Sector 15 (Camera 4; August 15, 2019 -- September 11, 2019), Sector 21 (Camera 3; January 21, 2020 -- February 18, 2020), and Sector 22 (Camera 3; February 18, 2020 -- March 18, 2020). TOI-1266 is listed as TIC 467179528 in the TESS Input Catalog \citep[TIC;][]{stassun2018,stassun2019}. Pixel data in a $11 \times 11$ array surrounding TOI-1266 were averaged into 2-minute stacks, which were reduced to lightcurves by the Science Processing Operations Center (SPOC) at NASA Ames \citep{jenkins2016}. We analyzed the Presearch Data Conditioning Single Aperture Photometry (PDCSAP) lightcurve, which contains systematics-corrected data using the algorithms originally developed for the \textit{Kepler} data analysis pipeline. The PDCSAP lightcurve uses pixels chosen to maximize the SNR of the target and has removed systematic variability by fitting out trends common to many stars \citep{smith2012,stumpe2014}. Figure \ref{fig:fov} highlights the TESS apertures for the different TESS Sectors and nearby stars detected by Gaia. From Figure \ref{fig:fov}, we see that two stars partially overlap the TESS apertures for TOI-1266 (Tmag=11.0) in some sectors: TIC 467179527 (Tmag=15.6; separation of $36\arcsec$), and TIC 467179526 (Tmag=18.338, separation of $36\arcsec$), both of which are significantly fainter ($\Delta$Tmag=4.6, and $\Delta$Tmag=7.3) than TOI-1266. The faintness and the separation of the two stars results in minimal dilution of the TESS light curve.

Analysis by the TESS Science Processing Operations Center identified two possible planetary signals, and human vetting of the data reports \citep{twicken2018,li2019} resulted in the announcement of planet candidates TOI-1266.01 and TOI-1266.02, available on the TESS alerts website\footnote{\url{https://tev.mit.edu/data/}}. The SPOC data validation reports \citep{twicken2018,liu2019} note no significant centroid offsets for either planet candidate.

\begin{figure*}[t!]
\begin{center}
\includegraphics[width=\textwidth]{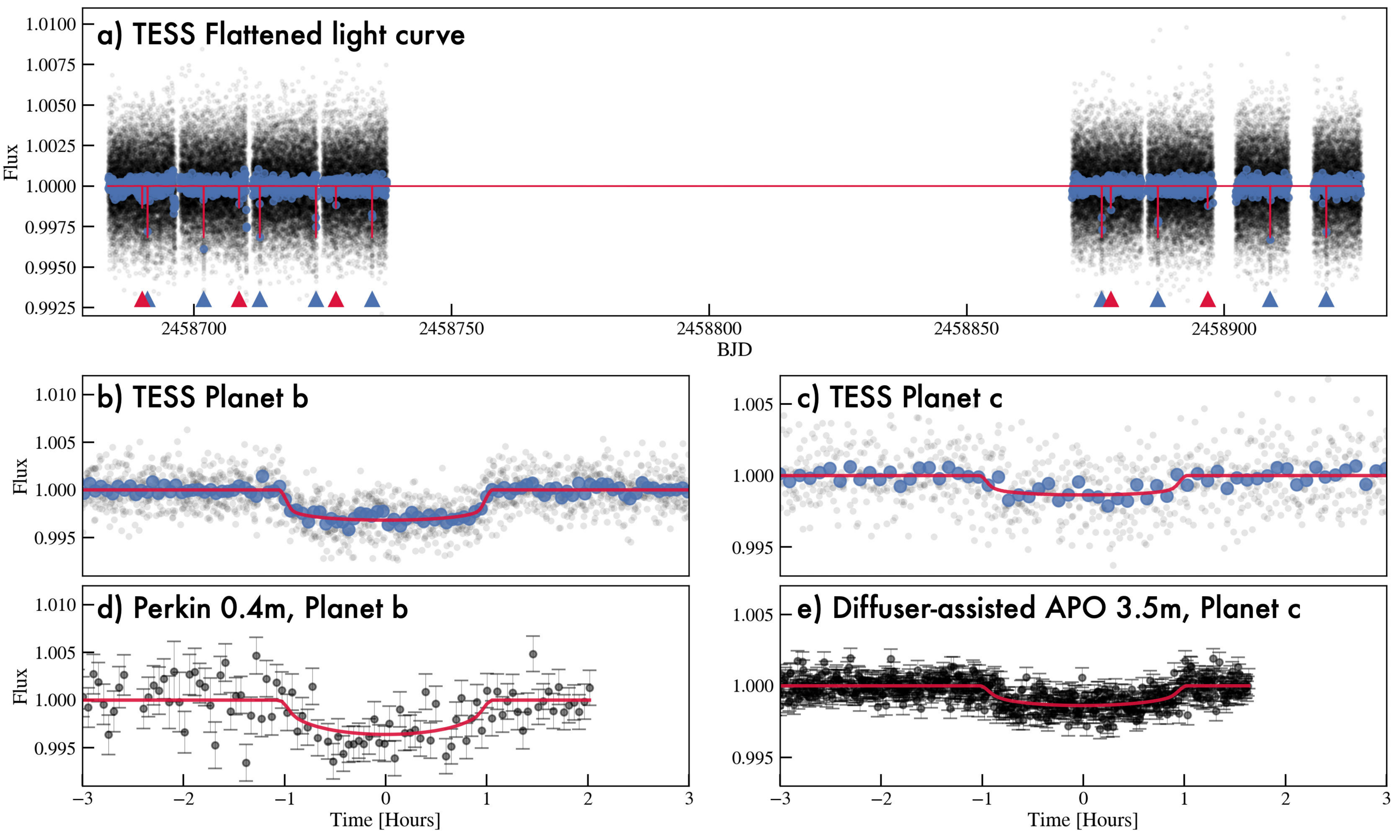}
\end{center}
\caption{Transit photometry of TOI-1266. a) Short-cadence (2-minute) TESS photometry is shown in black. The blue points show the data binned to 10 minutes. The red curve shows our best-fit joint model including both planets b and c. The blue and red triangles denote transits of planets b and c, respectively. b-c) Phase-folded photometry from TESS of the transits of TOI-1266b and TOI-1266c, respectively. d) Ground-based photometry from the 0.4m Perkin Telescope showing the transit of TOI-1266b. e) Diffuser-assisted photometry during the transit of TOI-1266c using the Engineered Diffuser on the 3.5m Telescope at Apache Point Observatory.}
\label{fig:transits}
\end{figure*}

To clean the available TESS data, we removed all points with non-zero quality flags (4844 in total) which indicate known problems \citep[e.g.,][]{tenenbaum2018}. We removed additional 12 points that we identified as 4 sigma outliers, leaving a total of 68891 points that we used for the fitting, with a median errorbar of 2270ppm. The median-normalized TESS $\mathrm{PDCSAP}$ light curve is shown in Figure \ref{fig:transits}. We retrieved the data using the \texttt{lightkurve} package \citep{lightkurve}.

\subsection{Ground-based Photometry with the 0.4m Perkin Telescope}
We observed a transit of TOI-1266b (Figure \ref{fig:transits}) on the night of March 21, 2020 using the 0.43m (17") Richard S. Perkin telescope at Hobart and William Smith Colleges. The telescope is a 17" PlaneWave Corrected Dall-Kirkham (CDK) telescope on a Paramount equatorial mount with an SBIG 8300 M camera with $3326 \times 2504$ pixels that are $5.4 \times 5.4 \micron$ square. The plate scale of the camera in the $1 \times 1$ binning mode we used is $0.38\unit{\arcsec/pixel}$, resulting in a Field-of-View (FOV) of $21 \times 16^\prime$. We obtained 106 images over $\sim$5 hours centered on the target in the Sloan $r^\prime$ filter, where all images were taken above an airmass of 1.5. To improve the observing efficiency, we defocused moderately, which allowed us to use an exposure time of 180 seconds. The guiding was stable throughout the observations. 

We processed the observations using AstroImageJ \citep{collins2017} using standard bias, dark, and flat-field frames. For flat-field calibrations, we used a median combined flat created from 28 sky-flat images at the beginning of the observations. We performed aperture photometry using AstroImageJ \citep{collins2017} on the calibrated images. We systematically tested a number of different apertures from 15 to 30 pixels. Ultimately, we settled on an aperture of 18 pixels ($6.8\arcsec$) in radius with inner and outer sky annuli of 35 pixels ($13.3\arcsec$) and 45 pixels ($17.1\arcsec$), respectively, which showed the lowest scatter in the final light curve. We experimented with detrending with different parameters (e.g., airmass, centroid offsets), but we observed no significant improvement in the resulting photometry.

\subsection{Diffuser-assisted Photometry with the 3.5m ARC Telescope}
We observed a transit of TOI-1266c (Figure \ref{fig:transits}) on the night of January 28, 2020 using the the Astrophysical Research Consortium Telescope Imaging Camera (ARCTIC) Imager \citep{huehnerhoff2016} on the 3.5m Astrophysical Research Consortium (ARC) 3.5m Telescope at Apache Point Observatory (APO). The target rose from an airmass of 1.44 at the start of the observations to a minimum airmass of 1.19, and ended at a slightly lower airmass of 1.21. We observed the transit using the Engineered Diffuser available on ARCTIC, which we designed specifically to enable precision photometric observations from the ground on nearby bright stars \citep[see e.g.,][]{stefansson2017,stefansson2018a,stefansson2018b,stefansson2020}. The observations were performed using the SDSS $i^\prime$ filter with an exposure time of $25 \unit{s}$ in the quad-readout mode with $4 \times 4$ on-chip binning. In this mode, ARCTIC has a gain of $2.0 \unit{e/ADU}$, and a plate scale of $0.44 \unit{\arcsec/pixel}$, and a short readout time of $2.7 \unit{s}$.

We processed the data using AstroImageJ \citep{collins2017} using standard bias and dark frames. We did observe a linear trend in the data, which through visual inspection could effectively be removed using a combination of detrending with a simultaneous line + airmass detrend. We experimented reducing the photometry both with and without a flat field calibration, but neither removed the observed trend. We saw a slight improvement in the resulting photometry without using the flat field, and as such, elected to present the data without the flat-field calibration. As discussed below, for our final parameter estimation, we fit for the transit model simultaneously with a Gaussian-Process model using a Matern 3/2 kernel to account for this red-noise component observed in the transit data. Clear outliers, either due to cosmic rays or charged-particle events were removed using AstroImageJ. To arrive at the final photometric reduction, we experimented extracting the data using a number of different apertures, and selected an aperture of 18 pixels ($8\arcsec$) with an inner sky annulus of 20 pixels ($9\arcsec$) and outer sky annulus of 50 pixels ($22\arcsec$), as this setting showed the overall lowest scatter in the final light curve.

\subsection{Habitable-zone Planet Finder}
We obtained high resolution spectra of TOI-1266 with the Habitable-zone Planet Finder (HPF) Spectrograph to place upper limits on the masses of both planets and to obtain precise spectroscopic parameters of the host star.  HPF is a fiber-fed near-infrared (NIR) spectrograph on the 10m Hobby-Eberly Telescope \citep{mahadevan2012,mahadevan2014} at McDonald Observatory in Texas, covering the $z$, $Y$, and $J$ bands from 810$\unit{nm}$-1260$\unit{nm}$ at a resolution of $R=55,000$. To enable precision radial velocities in the NIR, HPF is temperature stabilized at the milli-Kelvin level \citep{stefansson2016}. The HET is a fully queue scheduled telescope \citep{shetrone2007}, and all observations were executed as part of the HET queue. In total we obtained 46 spectra in 22 different HET tracks\footnote{HET is a fixed-altitude telescope and can only observe a given target at certain times or 'tracks'.} with two 969$\unit{s}$ exposures taken on average in each HET track. The 46 different spectra had a median SNR of 135 per extracted 1D pixel evaluated at 1~micron, and a median RV errorbar of 10.3$\unit{m/s}$. After binning to the 22 different individual tracks, the median RV errorbar is 7.4$\unit{m/s}$. We used the binned RVs for all subsequent analysis.

HPF has a NIR Laser Frequency Comb (LFC) calibrator to provide a precise wavelength solution and track instrumental drifts, which has been shown to enable $\sim$20$\unit{cm/s}$ RV calibration precision in 10 minute bins \citep{metcalf2019}. Following \cite{stefansson2020}, we elected not to use the simultaneous LFC calibration during the observations to minimize the risk of contaminating the science spectrum from scattered light from the LFC. Instead, we perform the RV drift correction by extrapolating the wavelength solution from LFC frames taken as part of standard evening/morning calibrations and from LFC calibration frames taken periodically throughout the night. This methodology has been shown to enable precise wavelength calibration at the $\sim$$30 \unit{cm/s}$ level, much smaller than the RV errorbar of the observations discussed here.

The HPF 1D spectra were reduced using the HPF pipeline, following the procedures in \cite{ninan2018}, \cite{kaplan2018}, and \cite{metcalf2019}. Following the 1D spectral extraction, we reduced the HPF radial velocities using an adopted version of the \texttt{SERVAL} (SpEctrum Radial Velocity Analyzer) pipeline \citep{zechmeister2018}, which is described in \cite{stefansson2020}. In short, \texttt{SERVAL} uses the template matching algorithm to derive RVs, which has been shown to be particularly effective at producing precise radial velocities for M-dwarfs \citep{anglada2012}. \texttt{SERVAL} uses the \texttt{barycorrpy} package \citep{kanodia2018} which uses the methodology of \cite{wright2014} to calculate accurate barycentric velocities. Following \cite{metcalf2019} and \cite{stefansson2020}, we only use the 8 HPF orders that are cleanest of tellurics, covering the wavelength regions from 8540-8890\AA, and 9940-10760\AA. We subtracted the estimated sky-background from the stellar spectrum using the dedicated HPF sky fiber. Again following the methodology described in \cite{metcalf2019} and \cite{stefansson2020}, we explicitly masked out telluric lines and sky-emission lines to minimize their impact on the RV determination. Table \ref{tab:rvs} in the Appendix lists the RVs from HPF used in this work.

\begin{figure}[t]
\begin{center}
\includegraphics[width=\columnwidth]{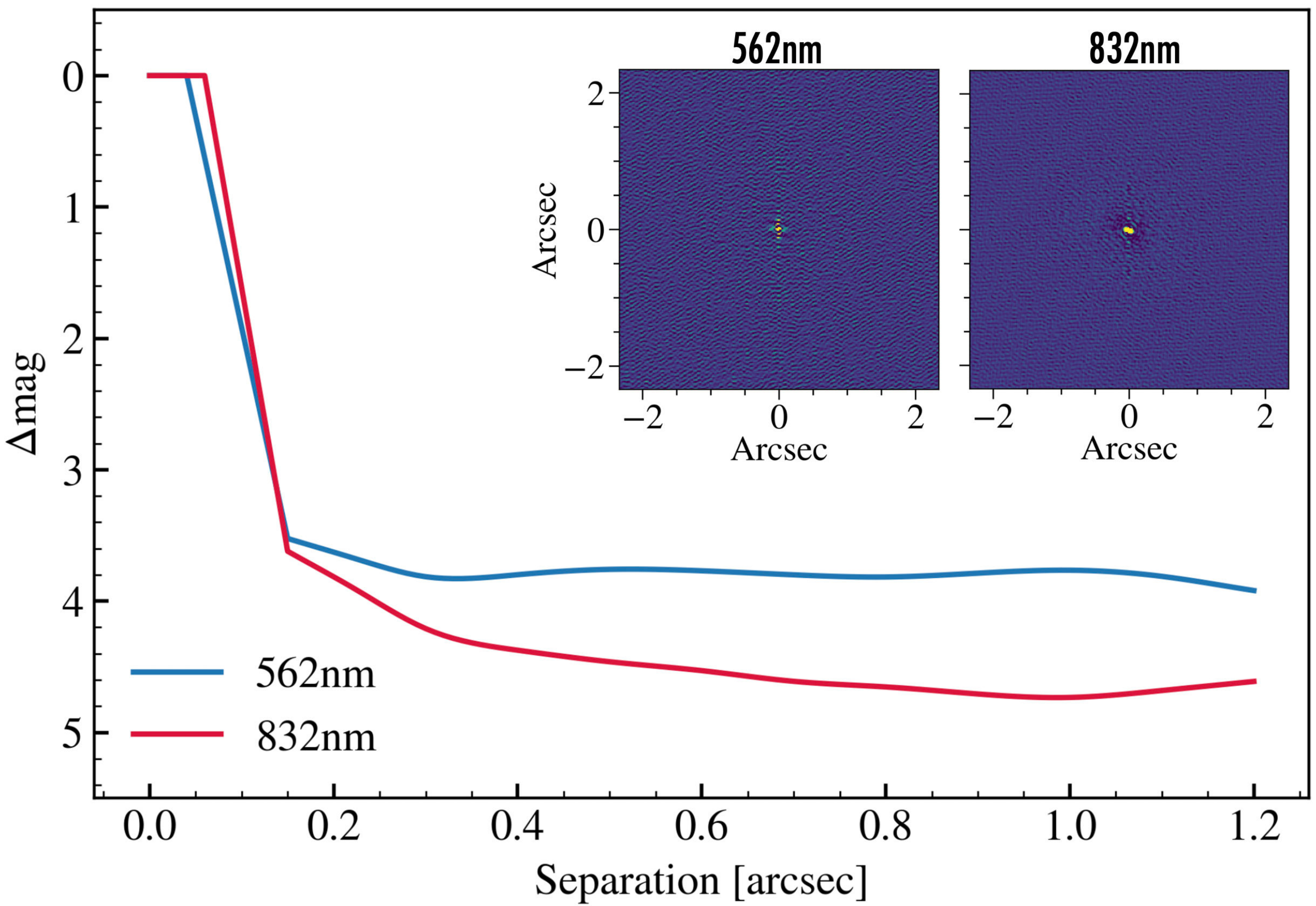}
\end{center}
\caption{Contrast limits from our NESSI speckle imaging data shown in two different bands centered around 562nm (blue) and 832nm (red). The insets show reconstructed images from the two bandpasses. No secondary sources are detected.}
\label{fig:nessi}
\end{figure}

\subsection{Speckle Imaging}
To rule out nearby companions, on the night of December 5 2019, we obtained speckle observations of TOI-1266 using the NASA Exoplanet Star and Speckle Imager \citep[NESSI;][]{scott2018} on the 3.5m WIYN Telescope at Kitt Peak National Observatory in Arizona. We reduced the data following the methodologies outlined in \cite{howell2011}. NESSI provides a resolution of $\sim$$0.04\arcsec$ in two bands centered around $562 \unit{nm}$ (width of $44 \unit{nm}$) and $832 \unit{nm}$ \citep[width of $40 \unit{nm}$;][]{scott2018}. Figure \ref{fig:nessi} shows the resulting contrast curves and reconstructed $256 \times 256$ images for the two bands. No secondary sources were detected in the reconstructed images, and from the contrast curve, we place a limit of $\Delta$mag$\sim$4 for nearby companions between $0.2\arcsec$ and $1.2\arcsec$.

\section{Stellar Parameters}
\label{sec:stellarparams}
To obtain spectroscopic constraints on the effective temperature $T_{\mathrm{eff}}$, stellar surface gravity $\log g$, and metallicity $[\mathrm{Fe/H}]$, we use the empirical spectral matching algorithm described in \cite{stefansson2020}. In short, this algorithm closely follows the methodology in \cite{yee2017}, where the target spectrum is compared to a library of high S/N as-observed spectra using a $\chi^2$ metric. From our analysis of the HPF spectra, we obtain the following spectroscopic values: $T_{\mathrm{eff}} = 3563 \pm 77 \unit{K}$, $\log g = 4.785 \pm 0.05$, $[\mathrm{Fe/H}] = -0.121 \pm 0.13$. From the spectral matching analysis, the two best matching stars are GJ 2066 and GJ 393, both of which have literature spectral types of M2.0 (see \citealt{floriano2015}, and \citealt{lepine2013}, respectively), which we adopt for TOI-1266.

To obtain model-dependent constraints on the stellar mass, radius, effective temperature, and age, we fit the Spectral Energy Distribution (SED) of TOI-1266 using the \texttt{EXOFASTv2} package \citep{eastman2019} using as inputs a) the available literature photometry, b) the Gaia distance from \cite{bailer-jones2018}, and c) the spectroscopic values discussed above as Gaussian priors. We adopt a uniform prior for the visual extinction where the upper limit is determined from estimates of Galactic dust by \cite{Green2019} (Bayestar19) calculated at the distance determined by \cite{bailer-jones2018}. We adopt the \(R_{v}=3.1\) reddening law from \cite{Fitzpatrick1999} to convert the Bayestar19 extinction to a visual magnitude extinction. \texttt{EXOFASTv2} uses the BT-NextGen Model grid of theoretical spectra \citep{Allard2012}, and the MESA Isochrones and Stellar Tracks \citep[MIST;][]{dotter2016,choi2016} to fit the SED and derive model dependent stellar parameters. Table \ref{tab:stellarparam} lists the resulting model dependent stellar parameters derived from the SED analysis, which agree well with the spectroscopically derived parameters. We calculate the galactic $U$, $V$, and $W$ velocities of TOI-1266 using the \texttt{GALPY} \citep{bovy2015} package (see Table \ref{tab:stellarparam}), and we note that \cite{carrillo2020} calculate membership probabilities of 97.2\%, 0.0\%, 2.8\% for TOI-1266 to be a member of the galactic thin-disk, thick-disk, and galactic halo populations, respectively.

From the spectral matching analysis we also obtain a limit on the projected stellar rotational velocity of $v \sin i < 2 \unit{km/s}$, suggestive of a slow rotator. This is in agreement with the the fact that we do not see clear rotational modulation in the TESS photometry at short periods. As a further test, we analyzed available ground-based photometry from the All-Sky Automated Survey for SuperNovae \citep[ASAS-SN;][]{kochanek2017} and the Zwicky Transient Facility \citep[ZTF;][]{masci2019}. We see no significant rotation signals that occur in both datasets by studying their Lomb-Scargle periodograms of these datasets. In addition, in Subsection \ref{sec:activity}, we discuss periodograms of activity indicators from the HPF spectra, which show no clear evidence of activity (e.g., no clear variability seen in the Calcium II Infrared Triplet or differential line widths). As such, without clear indication of photometric modulation in neither the TESS, ground-based photometry, or signs of activity from the HPF spectra, we conclude that TOI-1266 is an inactive star with a moderate or long rotation period.

\begin{deluxetable*}{llcc}
\tablecaption{Summary of stellar parameters used in this work. \label{tab:stellarparam}}
\tabletypesize{\scriptsize}
\tablehead{\colhead{~~~Parameter}                                 &  \colhead{Description}                                            & \colhead{Value}                         & \colhead{Reference}}
\startdata
\multicolumn{4}{l}{\hspace{-0.2cm} Main identifiers:}                                                                                                                                                     \\
TIC                                                               &  -                                                                & 467179528                               & TIC                     \\
TOI                                                               &  -                                                                & 1266                                    & TIC                     \\
2MASS                                                             &  -                                                                & J13115955+6550017                       & TIC                     \\
\multicolumn{4}{l}{\hspace{-0.2cm} Equatorial Coordinates, Proper Motion and Spectral Type:}           \\
$\alpha_{\mathrm{J2000}}$                                         &  Right Ascension (RA)                                             & 13:11:59.18                             & Gaia                    \\
$\delta_{\mathrm{J2000}}$                                         &  Declination (Dec)                                                & +65:50:01.31                            & Gaia                    \\
$\mu_{\alpha}$                                                    &  Proper motion (RA, \unit{mas\ yr^{-1}})                          & $-150.652 \pm 0.041$                    & Gaia                    \\
$\mu_{\delta}$                                                    &  Proper motion (Dec, \unit{mas\ yr^{-1}})                         &  $-25.368 \pm 0.039$                    & Gaia                    \\
Spectral Type                                                     &  -                                                                & M2                                      & This Work               \\
\multicolumn{4}{l}{\hspace{-0.2cm} Equatorial Coordinates, Proper Motion and Spectral Type:}           \\
$B$                                                               &  APASS Johnson B mag                                              & $14.578 \pm 0.048$                       & APASS                  \\
$V$                                                               &  APASS Johnson V mag                                              & $12.941 \pm 0.049$                       & APASS                  \\
$g^{\prime}$                                                      &  APASS Sloan $g^{\prime}$ mag                                     & $13.811 \pm 0.050$                       & APASS                  \\
$r^{\prime}$                                                      &  APASS Sloan $r^{\prime}$ mag                                     & $12.297 \pm 0.070$                       & APASS                  \\
$i^{\prime}$                                                      &  APASS Sloan $i^{\prime}$ mag                                     & $11.246 \pm 0.150$                       & APASS                  \\ 
\textit{TESS}-mag                                                 &  \textit{TESS} magnitude                                          & $11.040 \pm 0.007$                       & TIC                    \\
$J$                                                               &  2MASS $J$ mag                                                    & $9.706 \pm 0.023$                        & 2MASS                  \\
$H$                                                               &  2MASS $H$ mag                                                    & $9.065 \pm 0.030$                        & 2MASS                  \\
$K_S$                                                             &  2MASS $K_S$ mag                                                  & $8.840 \pm 0.020$                        & 2MASS                  \\
$WISE1$                                                           &  WISE1 mag                                                        & $8.715 \pm 0.022$                        & WISE                   \\
$WISE2$                                                           &  WISE2 mag                                                        & $8.612 \pm 0.019$                        & WISE                   \\
$WISE3$                                                           &  WISE3 mag                                                        & $8.504 \pm 0.024$                        & WISE                   \\
$WISE4$                                                           &  WISE4 mag                                                        & $8.233 \pm 0.207$                        & WISE                   \\
\multicolumn{4}{l}{\hspace{-0.2cm} Spectroscopic Parameters$^a$:}           \\
$T_{\mathrm{eff}}$                                                &  Effective temperature in \unit{K}                                & $3563 \pm 77$                            & This work              \\
$\mathrm{[Fe/H]}$                                                 &  Metallicity in dex                                               & $-0.121 \pm 0.13$                        & This work              \\
$\log(g)$                                                         &  Surface gravity in cgs units                                     & $4.785 \pm 0.05$                         & This work              \\
\multicolumn{4}{l}{\hspace{-0.2cm} Model-Dependent Stellar SED and Isochrone fit Parameters$^b$ (adopted):}           \\
$T_{\mathrm{eff}}$                                                &  Effective temperature in \unit{K}                                & $3573_{-38}^{+35}$                       & This work              \\
$\mathrm{[Fe/H]}$                                                 &  Metallicity in dex                                               & $-0.08_{-0.10}^{+0.13}$                  & This work              \\
$\log(g)$                                                         &  Surface gravity in cgs units                                     & $4.826_{-0.021}^{+0.020}$                & This work              \\
$M_*$                                                             &  Mass in $M_{\odot}$                                              & $0.437 \pm 0.021$                        & This work              \\
$R_*$                                                             &  Radius in $R_{\odot}$                                            & $0.4232_{-0.0079}^{+0.0077}$             & This work              \\
$\rho_*$                                                          &  Density in $\unit{g\:cm^{-3}}$                                   & $8.13_{-0.46}^{+0.47}$                   & This work              \\
Age                                                               &  Age in Gyrs                                                      & $7.9_{-5.2}^{+4.2}$                      & This work              \\
$L_*$                                                             &  Luminosity in $L_\odot$                                          & $0.02629_{-0.00075}^{+0.00071}$          & This work              \\
$A_v$                                                             &  Visual extinction in mag                                         & $0.015_{-0.010}^{+0.011}$                & This work              \\
$d$                                                               &  Distance in pc                                                   & $36.011_{-0.030}^{+0.029}$               & Gaia, Bailer-Jones     \\
$\pi$                                                             &  Parallax in mas                                                  & $27.769_{-0.022}^{+0.023}$               & Gaia                   \\
\multicolumn{4}{l}{\hspace{-0.2cm} Other Stellar Parameters:}           \\
$v \sin i_*$                                                      &  Stellar rotational velocity in \unit{km\ s^{-1}}                 &  $<2$                                    & This work              \\
$RV$                                                              &  Absolute radial velocity in \unit{km\ s^{-1}} ($\gamma$)         & $-41.58 \pm 0.26$                        & This work              \\
$U$.                                                              &  Galactic $U$ Velocity (km/s)                                     & $-5.8\pm0.2$                             & This work              \\
$V$                                                               &  Galactic $V$ Velocity (km/s)                                     & $-40.3\pm0.4$                            & This work              \\
$W$                                                               &  Galactic $W$ Velocity (km/s)                                     & $-27.9\pm0.6$                            & This work              \\
\enddata
\tablenotetext{}{References are: TIC \citep{stassun2018,stassun2019}, Gaia \citep{gaia2018}, APASS \citep{henden2015apass}, 2MASS/WISE \citep{cutri2014wise}, Bailer-Jones \citep{bailer-jones2018}.}
\tablenotetext{a}{Derived using the HPF spectral matching algorithm from \cite{stefansson2020}.}
\tablenotetext{b}{{\tt EXOFASTv2} derived values using MIST isochrones with the Gaia parallax and spectroscopic parameters in $a$) as priors.}
\end{deluxetable*}

\section{Planet Parameters}
\label{sec:planetparams}

\subsection{Search for Additional Planets}
\label{sec:activity}
\begin{figure}[t!]
\begin{center}
\includegraphics[width=\columnwidth]{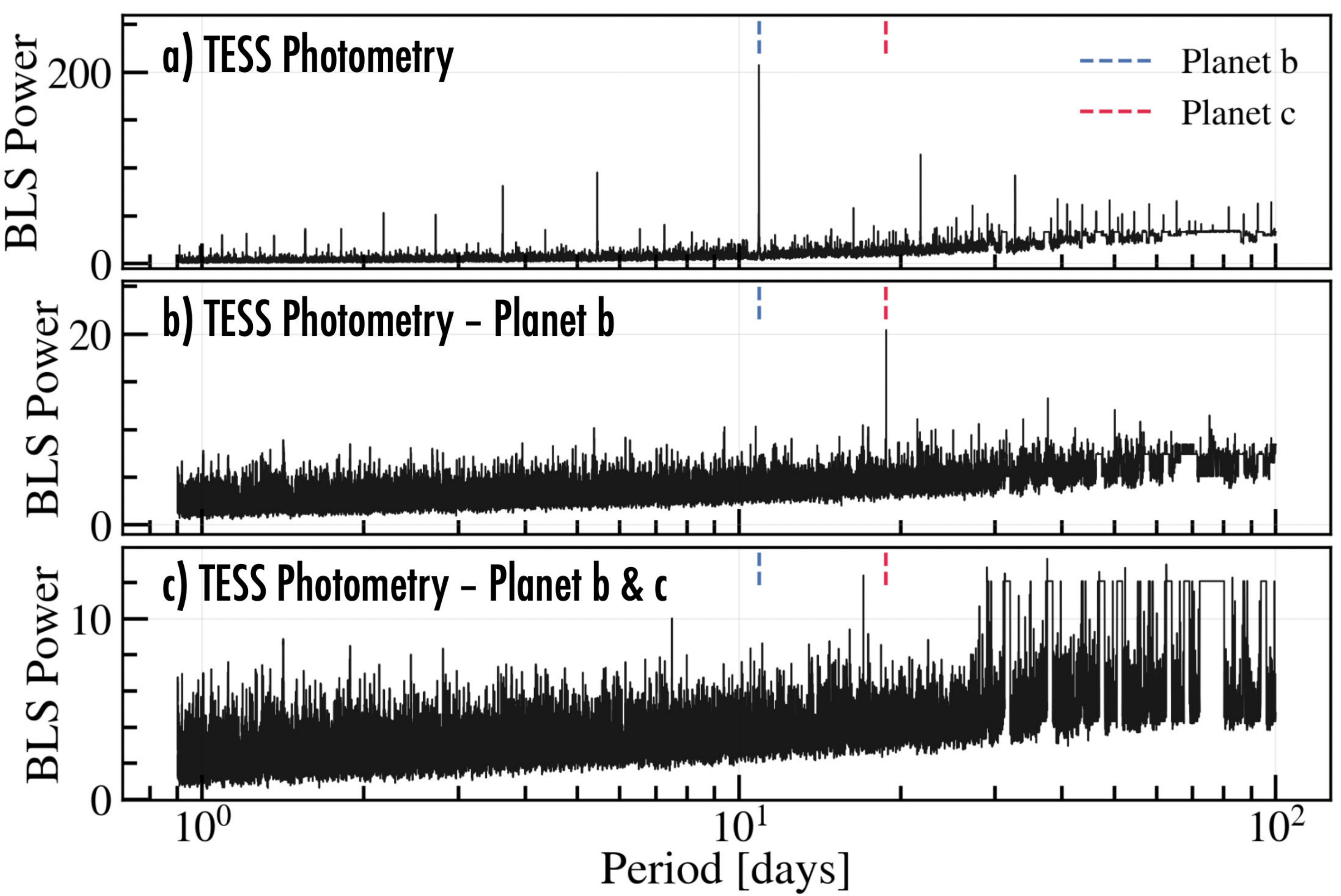}
\end{center}
\caption{Box-Least-Square (BLS) power spectra as a function of orbital period: a) BLS power spectrum of all available TESS photometry shows a clear peak at a period of $P=10.89$days (planet b, blue vertical line); b) BLS power spectrum of the TESS photometry after masking out transits of planet b shows a clear peak at $P=18.80$days (planet c, red vertical line); BLS power spectrum after masking out both transits of planet b and c shows no further clear peaks.}
\label{fig:bls}
\end{figure}

We looked for additional transiting planets in the TESS data using the Box-Least-Squares (BLS) algorithm \citep{kovacs2002} as implemented in the \texttt{lightkurve} package. Figure \ref{fig:bls} shows the BLS power spectra of the available TESS photometry after iteratively masking out transits of planets b and c (in a region 1.5 times as wide as the transit duration for each planet centered around the transit midpoints), showing no significant evidence for further transiting planets in the system. We additionally looked for evidence of Transit Timing Variations \citep[TTVs;][]{holman2005,agol2005} using the \texttt{TTVOrbit} fitting tools in the \texttt{exoplanet} code \citep{exoplanet2020}. In doing so, we see no evidence for significant TTVs, with all individual transit times fully consistent with a linear ephemeris, which suggests that there are no massive planets in the system orbiting at or close to orbital resonances with planets b or c.

\begin{figure}
\begin{center}
\includegraphics[width=1.0\columnwidth]{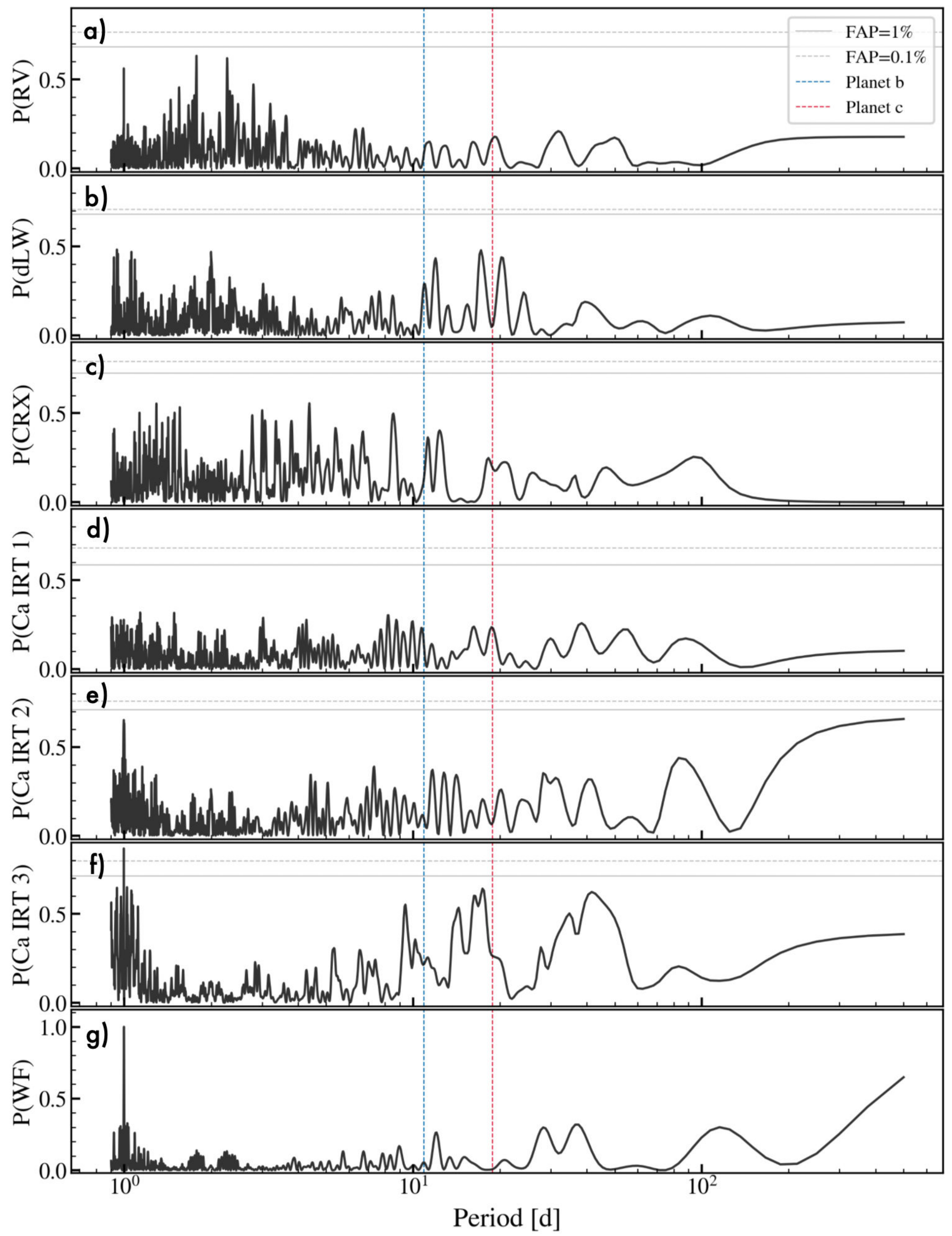}
\caption{Lomb-Scargle periodograms of the HPF RVs along with different activity indicators. The periods of planets b and c are highlighted with the dashed blue and red lines, respectively. False Alarm Probabilities (FAP) of 1\% and 0.1\% calculated using a bootstrap method are denoted with the grey solid and grey dashed lines, respectively. a) HPF RVs; b) Differential line width (dLW) activity indicator; c) Chromatic Index activity indicator (CRX); d-f) Ca II IRT indices for the three Ca II IRT lines; g) The window function of the HPF RVs, showing a clear sampling peak at 1 day. The power in a-f) is normalized using the formalism in \cite{zechmeister2009}, and g) is normalized so that the highest peak is unity.}
\label{fig:periodograms}
\end{center}
\end{figure}

We additionally looked for signs of non-transiting planets in the HPF RVs. Figure \ref{fig:periodograms} shows Generalized Lomb-Scargle (LS) periodograms of the HPF RVs, along with a number of activity indicators measured from the HPF spectra, including the Differential Line Width (dLW), the Chromatic Index (CRX), and line indices of the three Calcium II Infrared Triplet (Ca II IRT) lines. To calculate the activity indicators, we follow the definition and procedures in the \texttt{SERVAL} pipeline \citep{zechmeister2018}, and we note that their use for HPF spectra, including listing the exact wavelength ranges used to calculate the Ca II IRT indices, are further discussed in Stefansson et al. 2020 (submitted). We calculate the Generalized LS periodograms using the \texttt{astropy.timeseries} package, and we calculated the False Alarm Probabilities\footnote{Although the False Alarm Probability is a commonly used in periodogram analysis in radial velocity data it has known limitations \citep[see e.g., discussion in][]{fischer2016}.} using the \texttt{bootstrap} method implemented in the same package. In Figure \ref{fig:periodograms}, we additionally show the Window Function (WF) of our RV observations. All of the periodograms in Figure \ref{fig:periodograms} are normalized using the formalism in \cite{zechmeister2009}, except the window function is normalized such that the highest peak has a power of 1. Table \ref{tab:rvs} in the appendix lists the values of the RVs and the activity indicators.

From Figure \ref{fig:periodograms}, we see no significant peaks (with $\mathrm{FAP}<0.1\%$), with no clear peaks seen at the known planet periods. We attribute the latter due the expected RV amplitude of the planets (3.3m/s and 1.6m/s for planets b and c, respectively) being below the median HPF RV precision of 7.4m/s (see Subsection \ref{sec:mr}). We note that we see a hint of two peaks at 1.779days and its 1-day alias of 2.230days in the RVs (Figure \ref{fig:periodograms}a), although both peaks have a low significance with a $\mathrm{FAP}>1\%$. Although there remains a possibility that there are other planets in the system which could contribute additional variability to the RVs, further data is required to confidently rule out or confirm their presence. In the absence of strong evidence for more planets in the system, we fit the available datasets (photometry and RVs) assuming the two known transiting planets in the system.

\subsection{Transit, RV, and Gaussian Process Modeling}
We jointly model the available photometry from TESS and the two ground-based transits along with the radial velocities using the \texttt{juliet} code \citep{Espinoza2018}. In \texttt{juliet}, we used the \texttt{dynesty} package \citep{speagle2019} to perform dynamic nested sampling for parameter estimation. \texttt{juliet} uses the \texttt{batman} package \citep{kreidberg2015} for the transit model---which uses the transit prescription from \citep{mandel2002}---and uses the \texttt{radvel} package \citep{fulton2018} for the RV model. Following the implementation in \texttt{juliet}, we parameterize the transit in terms of the radius ratio ($p = \mathrm{R_p/R_*}$) and the impact parameter $b$. Due to the lack of nearby bright stars in the TESS aperture, and the resulting minimal dilution in the TESS data, we fix the dilution factor $D$ in \texttt{juliet} for the TESS and ground-based photometry to $D=0$. As both the ground-based and TESS transits were observed in similar band-passes (TESS bandpass, and in the SDSS $r^\prime$ and SDSS $i^\prime$ filters), we assume the transit depth in the TESS and ground-based transits are identical. We use a quadratic limb-darkening law to describe the transits, where we elect to use the $q_1$ and $q_2$ limb-darkening parameterization from \cite{kipping2013}, and to minimize biases in the resulting planet parameter constraints, we follow the suggestion in \cite{espinoza2015} and place uniform priors on the limb darkening parameters from 0 to 1.

To check if both transits recovered consistent stellar densities, we first performed a fit assuming circular orbits for both planets without an explicit prior on the stellar density. In doing so, we recover a stellar density of $\rho_* = 9.2 \pm 1.4 \unit{g/cm^3}$ and $\rho_* = 7.0_{-3.9}^{+5.0} \unit{g/cm^3}$ from the transits of planets b and c, respectively. From this, we see that both values are consistent with the model-dependent stellar density from Table \ref{tab:stellarparam} of $\rho_* = 8.13 \pm 0.48 \unit{g/cm^3}$, suggesting that the two planets indeed transit TOI-1266. This consistency between the transit-derived stellar density assuming circular orbits and the model-dependent stellar density further suggests that both planets have low eccentricities, which conforms with the trend that multi-transiting systems generally show low eccentricities \citep{vaneylen2015}. As such, without strong evidence suggesting non-circular orbits, for our final parameter estimation, we assumed that both planets have circular orbits. We further place a Gaussian prior on the stellar density of $\rho_* = 8.13 \pm 0.48 \unit{g/cm^3}$ to accurately constrain the orbital distance ($a/R_*$) of both planets. In total, we fit for 36 parameters. Table \ref{tab:priors} summarizes the priors we used.

\begin{deluxetable}{llc}
\tabletypesize{\scriptsize}
\tablecaption{Summary of priors used for our joint transit and RV fit. $\mathcal{N}(\mu,\sigma)$ denotes a normal prior with mean $\mu$, and standard deviation $\sigma$; $\mathcal{U}(a,b)$ denotes a uniform prior with a start value $a$ and end value $b$, and $\mathcal{J}(a,b)$ denotes a Jeffreys prior truncated between a start value $a$ and end value $b$. We assumed circular orbits and no photometric dilution for both planets.\label{tab:priors}}
\tablehead{\colhead{Parameter}&  \colhead{Description}                             & \colhead{Model}} 
\startdata	
\hline
\multicolumn{3}{l}{\hspace{-0.3cm} Orbital Parameters - Planet b:}                       \\
$P$                                                                           &  Orbital Period (days)                   & $\mathcal{N}(10.895411,0.01)$        \\
$T_C$                                                                         &  Transit Midpoint - 2458000 $(\mathrm{BJD_{TDB}})$ & $\mathcal{U}(690.95,691.05)$ \\
$R_{p}/R_{*}$                                                                 &  Scaled Radius                           & $\mathcal{U}(0,1)$                   \\
$a/R_*$                                                                       &  Scaled Semi-major axis                  & $\mathcal{J}(1,200)$                 \\
$b$                                                                           &  Impact Parameter                        & $\mathcal{U}(0,1)$                   \\
$K$                                                                           &  RV semi-amplitude ($\unit{m/s}$)        & $\mathcal{U}(0,100)$                 \\
\multicolumn{3}{l}{\hspace{-0.3cm} Orbital Parameters - Planet c:}                       \\
$P$                                                                           &  Orbital Period (days)                   & $\mathcal{N}(18.79545,0.01)$        \\
$T_C$                                                                         &  Transit Midpoint - 2458000 $(\mathrm{BJD_{TDB}})$ & $\mathcal{U}(689.90,690.00)$ \\
$R_{p}/R_{*}$                                                                 &  Scaled Radius                           & $\mathcal{U}(0,1)$                   \\
$a/R_*$                                                                       &  Scaled Semi-major axis                  & $\mathcal{J}(1,200)$                 \\
$b$                                                                           &  Impact Parameter                        & $\mathcal{U}(0,1)$                   \\
$K$                                                                           &  RV semi-amplitude ($\unit{m/s}$)        & $\mathcal{U}(0,100)$                 \\
\multicolumn{3}{l}{\hspace{-0.3cm} Other constraints:}                        \\
$\rho_*$                                                                      &  Stellar density ($\unit{g\:cm^{-3}}$)   & $\mathcal{N}(8.13,0.48)$             \\
\multicolumn{3}{l}{\hspace{-0.3cm} Instrumental Terms:}      \\
$q_1^{a}$                                                                     &  Limb-darkening parameter                & $\mathcal{U}(0,1)$                   \\
$q_2^{a}$                                                                     &  Limb-darkening parameter                & $\mathcal{U}(0,1)$                   \\
$\sigma_{\mathrm{phot}}$$^b$                                                  &  Photometric jitter ($\unit{ppm}$)       & $\mathcal{J}(1,5000)$                \\
$\mu_{\mathrm{phot}}$$^b$                                                     &  Photometric baseline                    & $\mathcal{N}(0,0.1)$                 \\
$\sigma_{\mathrm{HPF}}$                                                       &  HPF RV jitter (m/s)                     & $\mathcal{J}(0.01,100)$              \\
$\gamma$                                                                      &  HPF RV offset (m/s)                     & $\mathcal{U}(-50,50)$                \\
\multicolumn{3}{l}{\hspace{-0.3cm} TESS Quasi-Periodic GP Parameters:} \\
$P_{\mathrm{GP}}$                                                             &  GP Period (days)                        & $\mathcal{J}(0.1,1000)$              \\
$B$                                                                           &  GP Amplitude ($\unit{ppm^2}$)           & $\mathcal{J}(10^{-6},1)$             \\
$C$                                                                           &  GP Additive Factor                      & $\mathcal{J}(10^{-3},10^{3})$        \\
$L$                                                                           &  GP Length Scale (days)                  & $\mathcal{J}(1, 10^{3})$             \\
\multicolumn{3}{l}{\hspace{-0.3cm} Perkin 0.4m Approximate Matern 3/2 GP Parameters:} \\
$\sigma_{\mathrm{GP}}$                                                        &  GP Amplitude (ppm)                      & $\mathcal{J}(0.1, 10^4)$             \\
$\tau$                                                                        &  Timescale of exp. kernel (days)         & $\mathcal{J}(0.01, 10^{5})$          \\
$\rho$                                                                        &  Timescale of Matern kernel (days)       & $\mathcal{J}(0.01, 10^{5})$          \\
\multicolumn{3}{l}{\hspace{-0.3cm} APO 3.5m Approximate Matern 3/2 GP Parameters:} \\
$\sigma_{\mathrm{GP}}$                                                        &  GP Amplitude (ppm)                      & $\mathcal{J}(0.1, 10^4)$             \\
$\tau$                                                                        &  Timescale of exp. kernel (days)         & $\mathcal{J}(0.01, 10^{5})$          \\
$\rho$                                                                        &  Timescale of Matern kernel (days)       & $\mathcal{J}(0.01, 10^{5})$          \\
\enddata
\tablenotetext{a}{We use the same uniform priors for pairs of limb darkening parameters $q_1$ and $q_2$ (parameterization from \cite{kipping2013}, and use separate limb darkening parameters for each instrument).}
\tablenotetext{b}{For each photometric dataset (TESS, 0.4m Perkin, and 3.5m APO), we placed a separate photometric jitter and and baseline offset term.}
\end{deluxetable}

To account for correlated noise in the photometric datasets, we use a Gaussian Process noise model, where we choose different kernels for the different datasets to best reflect the characteristic noise structures seen in the data as a function of time. For the TESS data, to account for any possible low-level photometric modulations, we use the quasi-periodic kernel from the \texttt{celerite} package \citep{Foreman-Mackey2017}, with a kernel function of the form,
\begin{equation}
k(x_l,x_m) = \frac{B}{2 + C} e^{-\tau/L} \left[ \cos \left( \frac{2 \pi \tau}{P_\mathrm{GP}} \right) + (1 + C) \right],
\label{eq:kernelperiodic}
\end{equation}
where $\tau = |x_l - x_m|$, and where $B$, $C$, $L$, and $P_{\mathrm{rot}}$ are the hyperparameters of the kernel. $B$ and $C$ tune the weight of the exponential decay component of the kernel with a decay constant of $L$ (in days), and $P_{\mathrm{GP}}$ corresponds to the periodicity of the quasi-periodic oscillations which we interpret as the stellar rotation period. For the ground-based datasets, we follow Stefansson et al.~2020 (submitted), and use the Approximate Matern-3/2 kernel multiplied by an exponential kernel available in \texttt{juliet}. This kernel has covariance properties that are better matched to shorter-term instrumental and/or atmospheric red-noise structures often seen in ground-based datasets \citep[see e.g.,][]{pepper2017,Espinoza2018}. As implemented in \texttt{juliet}, this kernel has the following form (see also \cite{Foreman-Mackey2017}),
\begin{equation}
k(x_l,x_m) = \sigma_{\mathrm{GP}}^2 e^{-\tau/L} \left[ (1 + 1/\epsilon) e^{-1(1-\epsilon)s} + (1 - 1/\epsilon) e^{-1(1+\epsilon)s} \right],
\label{eq:kernelmat32}
\end{equation}
where $s = \sqrt{3} \tau / \rho$, and $\tau = |x_l - x_m|$, with hyperparameters $\sigma_{\mathrm{GP}}$ (photometric amplitude in $\mathrm{ppm}$), $L$ (length scale of the exponential component in days), and $\rho$ (length scale of the Matern-3/2 kernel in days), and with $\epsilon = 0.01$, where we note that as $\epsilon$ approaches 0 the factor inside the brackets converges to a Matern 3/2 kernel \citep{Foreman-Mackey2017,Espinoza2018}. 

For the RV dataset, given the few number of RV points available and the low activity of the star, we do not use a Gaussian Process model and rather adopt a white-noise model to account for potential systematics and/or stellar jitter effects.

\subsection{Derived Planet Parameters}
\begin{figure*}[t!]
\begin{center}
\includegraphics[width=\textwidth]{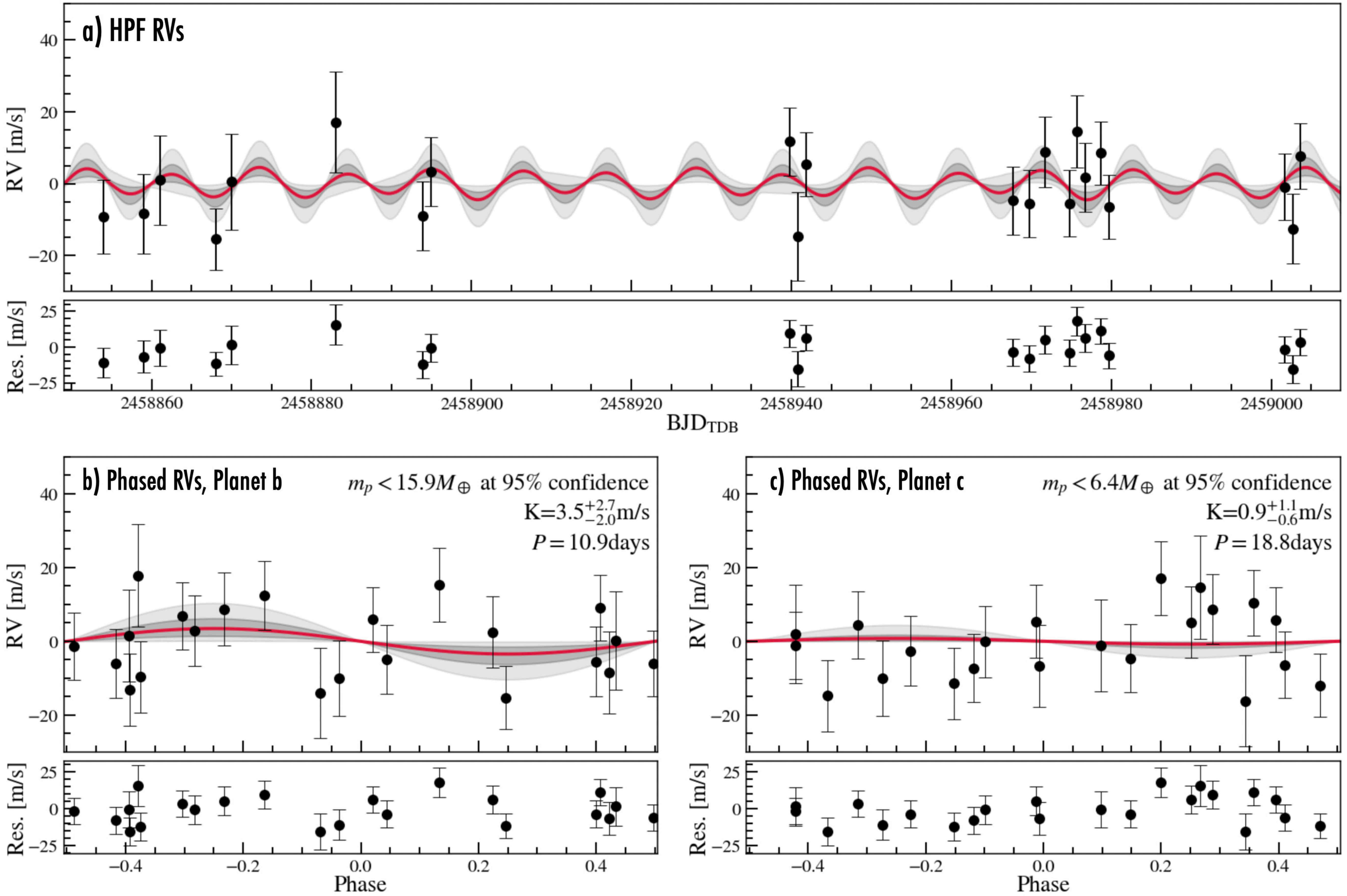}
\end{center}
\caption{RVs from HPF a) as a function of time, and b) and c) show the RVs folded on the periods of planet b and planet c, respectively. The median best-fit model is shown in red. The grey shaded regions show the 68\% and 99.7\% credible intervals from the posteriors.}
\label{fig:rvs}
\end{figure*}

Figure \ref{fig:transits} shows the TESS transits and ground-based transits, along with our best-fit model. Figure \ref{fig:rvs} shows the RVs from HPF, showing the unbinned RVs as a function of time, as well as the RVs phased around each planet. Table \ref{tab:planetparams} shows the resulting planet parameters from our joint fit of the photometry and the radial velocities. To cross-check the parameters reported by \texttt{juliet} which uses nested sampling, we performed a separate fit using the \texttt{exoplanet} code \citep{exoplanet2020}, which uses the \texttt{PyMC3} Markov Chain Monte-Carlo package for parameter estimation \citep{exoplanet:pymc3}. The \texttt{exoplanet} package builds on the \texttt{theano} package \citep{exoplanet:theano} for the numerical infrastructure and uses the \texttt{starry} package \citep{luger2019} for the light-curve generation. This test resulted in fully consistent parameters (within $1\sigma$) to the parameters reported by \texttt{juliet}. For brevity, we adapt the parameters from \texttt{juliet} in Table \ref{tab:planetparams}.

\begin{deluxetable*}{llcccc}
\tablecaption{Median values and 68\% credible intervals from our joint fit of the photometry and radial velocity data of TOI-1266. Both planets are assumed to be on circular orbits. The formal 68\% credible interval for the masses of the two planets are $6.9_{-4.0}^{+5.5} M_\oplus$ and $1.9_{-1.3}^{+2.3} M_\oplus$ for planets b and c, which we use to place 95\% upper limits on the mass as listed below. \label{tab:planetparams}}
\vspace{-0.2cm}
\tablehead{                  \colhead{~~~Parameter}&                      \colhead{Description}  &       \colhead{Planet b}              & \colhead{Planet c}}
\startdata
                     $T_{C}$ $(\mathrm{BJD_{TDB}})$ &                            Transit Midpoint &    $2458691.005_{-0.0011}^{+0.0011}$ &  $2458689.9589_{-0.0050}^{+0.0060}$ \\
                                         $P$ (days) &                              Orbital period &    $10.894879_{-0.00007}^{+0.00007}$ &    $18.80152_{-0.00067}^{+0.00054}$ \\
                                          $R_p/R_*$ &                                Radius ratio &         $0.0532_{-0.0012}^{+0.0015}$ &        $0.0363_{-0.0022}^{+0.0017}$ \\
                                 $R_{p} (R_\oplus)$ &                 Planet radius (Earth radii) &            $2.458_{-0.073}^{+0.083}$ &           $1.673_{-0.110}^{+0.087}$ \\
                           $R_{p} (R_{\mathrm{J}})$ &               Planet radius (Jupiter radii) &         $0.2193_{-0.0066}^{+0.0074}$ &        $0.1492_{-0.0095}^{+0.0077}$ \\
			              $\delta_{p, \mathrm{K2}}$ &                               Transit depth &      $0.00283_{-0.00013}^{+0.00016}$ &     $0.00132_{-0.00016}^{+0.00013}$ \\
                                            $a/R_*$ &                   Normalized orbital radius &                 $37.9_{-3.5}^{+2.2}$ &             $52.66_{-0.73}^{+0.97}$ \\
                                           $a$ (AU) &    Semi-major axis (from $a/R_*$ and $R_*$) &         $0.0745_{-0.0069}^{+0.0046}$ &        $0.1037_{-0.0025}^{+0.0026}$ \\
  $\rho_{\mathrm{*,transit}}$ ($\mathrm{g/cm^{3}}$) &                             Density of star &                  $8.7_{-2.2}^{+1.6}$ &              $7.81_{-0.32}^{+0.44}$ \\
                                   $i$ $(^{\circ})$ &                         Transit inclination &              $89.36_{-0.33}^{+0.20}$ &          $89.225_{-0.043}^{+0.060}$ \\
                                                $b$ &                            Impact parameter &               $0.43_{-0.12}^{+0.16}$ &           $0.714_{-0.050}^{+0.035}$ \\
                                                $e$ &                               Eccentricity  &                                            \multicolumn{2}{c}{0 (adopted)} \\ 
                              $\omega$ ($^{\circ}$) &                      Argument of periastron &                                           \multicolumn{2}{c}{90 (adopted)} \\               
                              $T_{\mathrm{eq}}$ (K) &        Equilibrium temp. (assuming $a=0.3$) &              $410.0_{-15.0}^{+21.0}$ &               $347.1_{-8.0}^{+7.9}$ \\
                              $T_{\mathrm{eq}}$ (K) &        Equilibrium temp. (assuming $a=0.0$) &              $287.0_{-11.0}^{+15.0}$ &               $243.0_{-5.6}^{+5.6}$ \\
                                 $S$ ($S_{\oplus}$) &                             Insolation Flux &                $4.72_{-0.66}^{+1.0}$ &              $2.42_{-0.22}^{+0.23}$ \\
                                    $T_{14}$ (days) &                            Transit duration &         $0.0879_{-0.0016}^{+0.0017}$ &        $0.0853_{-0.0036}^{+0.0046}$ \\
                                    $T_{23}$ (days) &                            Transit duration &         $0.0767_{-0.0021}^{+0.0019}$ &        $0.0735_{-0.0045}^{+0.0056}$ \\
                                      $\tau$ (days) &                     Ingress/egress duration &       $0.00537_{-0.00060}^{+0.0014}$ &     $0.00589_{-0.00064}^{+0.00053}$ \\
                                          $K$ (m/s) &                           RV Semi-amplitude &                  $3.5_{-2.0}^{+2.7}$ &               $0.8_{-0.53}^{+0.97}$ \\
                                 $m_p$ ($M_\oplus$) &                                Planet mass  &          $<15.9$ at 95\% confidence &          $<6.4$ at 95\% confidence  \\
		            $\sigma_{\mathrm{w,HPF}}$ (m/s) &                              HPF RV jitter  &                                    \multicolumn{2}{c}{$6.5_{-1.6}^{+2.0}$} \\ 
	                                 $\gamma$ (m/s) &                              HPF RV offset  &                                    \multicolumn{2}{c}{$0.3_{-1.9}^{+1.8}$} \\
\enddata
\end{deluxetable*}

\section{Statistical Validation}
\label{sec:validation}
To estimate the probability that the transits we observed were due to astrophysical false positives, we used the statistical techniques of \cite{morton2012vespa} implemented in the Validation of Exoplanet Signals using a Probabilistic Algorithm (\texttt{VESPA}) package \citep{morton2015vespa}. \texttt{VESPA} calculates the false positive probability (FPP) of transiting planet candidates by simulating and determining the likelihood of a range of astrophysical false positive scenarios that could replicate the observed light curves, including background eclipsing binaries, eclipsing binaries, and hierarchical eclipsing binaries. As inputs to \texttt{VESPA}, we used a) the phase-folded TESS transit in a 2x transit duration window around the center of each transit, b) the position of the target in the sky, c) the 2MASS $J$, $H$, $K$, SDSS $g^\prime$, $r^\prime$, $i^\prime$, and TESS magnitudes, d) the Gaia parallax, e) the host star stellar effective temperature, surface gravity, and metallicity, and f) the maximum visual extinction from estimates of Galactic dust extinction \citep{Green2019}. These values are listed in Table \ref{tab:stellarparam}. 

In addition to the inputs above, \texttt{VESPA} requires two additional constraints. First, as we have ground-based transit observations of both planets recovering fully consistent transits with the TESS transits but at a finer pixel scale, we set the maximum separation for a background eclipsing object equal to the aperture radius used for the ground-based photometric extractions for planet b ($7\arcsec$ from Perkin) and c ($8\arcsec$ from APO), respectively. Second, we set the maximum depth of secondary eclipse equal to the RMS of the unbinned TESS lightcurve (2262~ppm). Assuming the more conservative approach that the transits of planets b and c are independent, we obtain a FPP rate of $8\times10^{-6}$ and $1.9\times10^{-3}$ for planets b and c, respectively. Although already showing low FPP values, we argue that the real false positive probabilities are even lower accounting for that false positive scenarios are less likely in multi-planet systems \citep[e.g.,][]{latham2011,lissauer2012}. We consider both planets statistically validated.

\section{Discussion}
\label{sec:discussion}

\subsection{Mass and Bulk Composition Constraints}
\label{sec:mr}
From the HPF RVs, we obtain formal mass constraints of $6.9_{-4.0}^{+5.5} M_\oplus$ and $1.9_{-1.3}^{+2.3} M_\oplus$ for planets b and c, respectively, which we use to place upper mass constraints of $15.9 M_\oplus$ and $6.4 M_\oplus$ at 95\% confidence ($2\sigma$) for the two planets, respectively. The corresponding 99.7\% percentile constraints are $22.3M_\oplus$ and $11.3M_\oplus$, respectively. We compared these mass constraints with the predicted masses calculated with the mass-radius relations in the \texttt{Forecaster} \citep{chen2017} and the \texttt{MRExo} \citep{kanodia2019} mass-radius packages. \texttt{Forecaster} uses a broken power-law mass-radius relation to predict exoplanet masses from their radii derived from a sample of exoplanets across different spectral types, while the \texttt{MRExo} package uses a non-parametric relation specifically trained on current M-dwarf planet systems with well measured masses and radii \citep{kanodia2019}. From \texttt{Forecaster}, we predict a mass of $6.6_{-2.8}^{+5.0} M_\oplus$ and $3.8_{-1.4}^{+2.6}M_\oplus$ for planets b and c, translating to expected RV semi-amplitudes of $3.3_{-1.4}^{+2.5} \unit{m/s}$, and $1.6_{-0.6}^{+1.0} \unit{m/s}$, respectively. From \texttt{MRExo}, we predict masses of $6.2^{+6.7}_{-3.2} M_\oplus$ and $2.9^{+5.1}_{-1.7}M_\oplus$ for planets b and c, respectively. We see that our current mass constraints are fully consistent with the predicted mass estimates.

\begin{figure}[t!]
\begin{center}
\includegraphics[width=1.0\columnwidth]{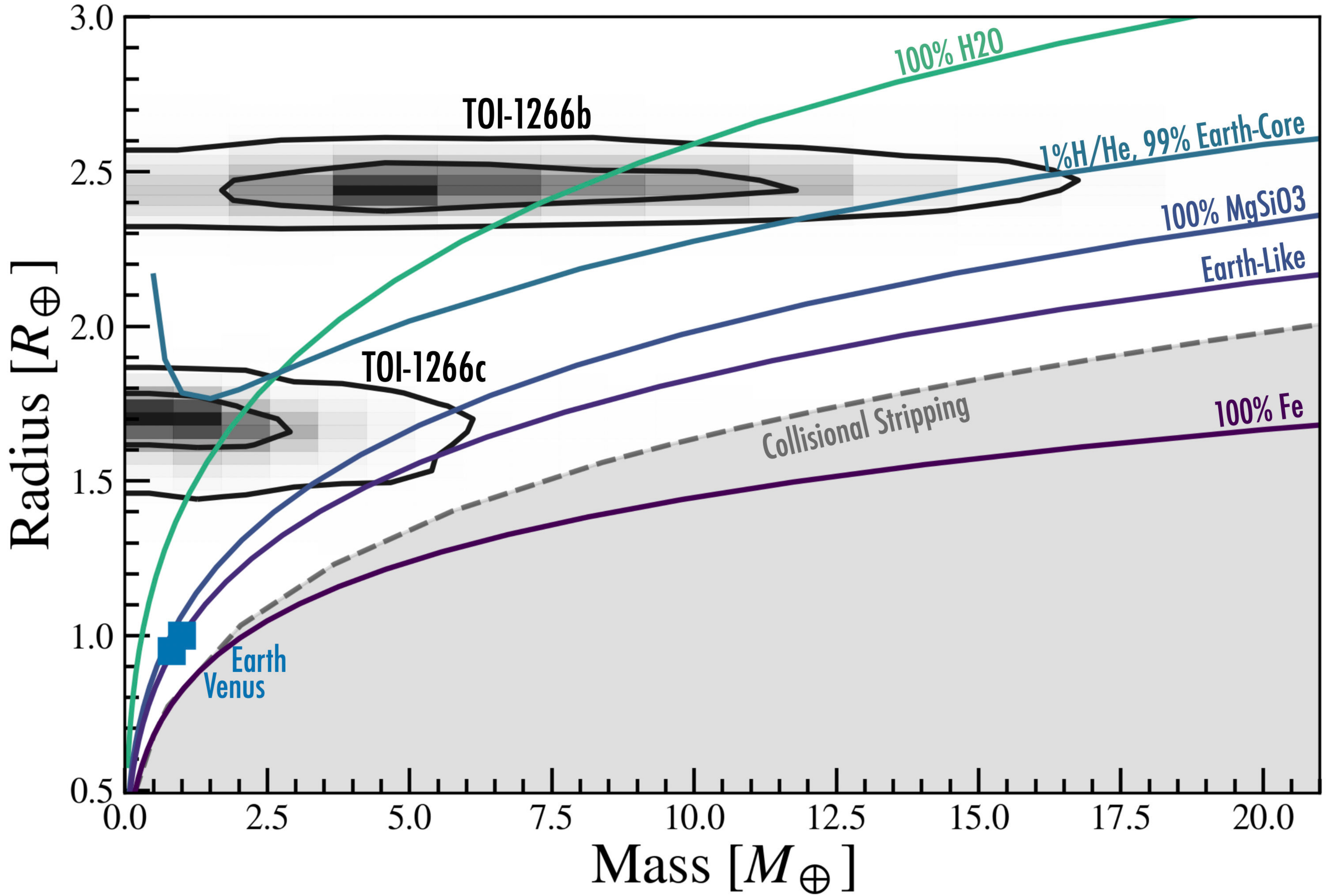}
\end{center}
\caption{Current radius and mass constraints of TOI-1266b and c from our joint 2-planet fit. The contours show our $1\sigma$ and $2\sigma$ posterior contours for planets b and c. The shaded grey region indicates planets with iron content exceeding the maximum value predicted from models of collisional stripping \citep{marcus2010}. The solid lines show different composition models from \cite{zeng2019}. Earth and Venus are denoted by blue squares. Further RV observations are needed to more precisely constrain the masses of both planets.}
\label{fig:mr}
\end{figure}

Using our formal mass constraints, in Figure \ref{fig:mr}, we explored the most likely composition of the two planets by comparing our posteriors to the composition models of \cite{zeng2019}. From Figure \ref{fig:mr}, we can see that both planets are consistent with non-rocky compositions, favoring either a water-rich world (e.g., the 100\% H$_{2}$O model) and/or a rocky core enveloped by a H/He atmosphere\footnote{In general, from exoplanet masses and radii alone we can not discern between such solutions, as there are degeneracies in the composition models of small planets \citep[see e.g.,][]{adams2008,zeng2019}.}. For planet b, if we assume the two-component model of \cite{lopez2014} consisting of a rocky core enveloped by a predominantly H/He atmosphere, we estimate a gas composition mass fraction of $1.5-2.0\%$. For planet c, although our current RVs currently only show a marginal non-zero detection of the low-RV amplitude signal, our current RV constraints suggesting a mass $<6.4 M_\oplus$ at 95\% confidence hint at a non-Earth-like composition, tilting towards a water-rich or rocky world enshrouded by a H/He atmosphere. Further RVs are required to confirm and better constrain the composition of both planets---in particular for planet c. As TOI-1266 is a relatively nearby and bright ($V=12.9$, $J=9.7$) early M-dwarf, an accurate mass measurement of both planets is within reach of current high precision spectrographs. To estimate the number of additional visits needed to measure the masses for transiting planets with known periods, we used the methodology of \cite{plavchan2015}. Assuming the RV semi-amplitudes expected from \texttt{Forecaster}, we estimate that we would need 10-20 more HPF visits to measure the mass of planet b at 99.7\% confidence ($3\sigma$), but measuring the mass of planet c is currently infeasible in <100 visits with HPF. However, as TOI-1266 is a relatively bright early M-dwarf, the RV information content is better matched for red-optical Doppler spectrographs such as NEID \citep{schwab2016}, CARMENES \citep{quirrenbach2018}, ESPRESSO \citep{pepe2018}, KPF \citep{gibson2016}, or MAROON-X \citep{seifahrt2016}. With NEID, assuming a 2.8m/s RV precision in 30minute bins, we estimate to be able to measure the masses of planets b and c in $\sim$6 and $\sim$30 visits at $3\sigma$, respectively.

\subsection{TOI-1266c Resides in the Radius Valley}
Close-in exoplanets display a gap or a valley in the radius distribution around 1.5-2.0 Earth radii \citep{owen2013,fulton2017,vaneylen2018,cloutier2020}, which has been interpreted as the transition radius between rocky and gaseous planets. A number of theoretical models have arisen to explain the emergence of the Radius Valley, which predict that the location of the rocky-to-gaseous transition radius, $r_{\mathrm{transition}}$, depends on the orbital period of the planet. The photoevaporation model, where the atmosphere of small planets can be stripped by high energy XUV photons leaving behind bare planetary cores \citep{lopez2012,owen2013,lopez2013,owen2017,lopez2018}, predicts that the transition radius should \textit{decrease} with orbital period as $r_{\mathrm{transition}} \propto P^{-0.15}$. Second, the core-powered mass-loss mechanism \citep{ginzburg2016,ginzburg2018,gupta2019}, where the luminosity of the cooling planetary core provides the energy for atmospheric loss, predicts that the transition radius should also \textit{decrease} with orbital period as $r_{\mathrm{transition}} \propto P^{-0.13}$. Third, in the gas-poor formation scenario, where super-Earths represent a distinct population of planets forming in a gas-poor environment after the protoplanetary disk has dissipated \citep{lee2014,lee2016,lopez2018}, the prediction is instead that the transition radius should \textit{increase} with orbital period as $r_{\mathrm{transition}} \propto P^{0.11}$.

To distinguish between these scenarios, previous studies have empirically measured the location of the Radius Valley as a function of orbital period. \cite{martinez2019} used data from \textit{Kepler} and the California Kepler Survey (CKS) to show that the location of the Radius Valley decreases as $r_{\mathrm{transition,M19}} \propto P^{-0.11 \pm 0.03}$ around Solar-type stars, consistent with mechanisms of photoevaporation and core-powered mass loss. This is in good agreement with the dependence of $r_{\mathrm{transition,VE18}} \propto P^{-0.09_{-0.04}^{+0.02}}$ measured by \cite{vaneylen2018} using a sample of planets orbiting solar-type stars with accurately determined stellar parameters from asteroseismology. Recently, \cite{cloutier2020} constrained the location of the Radius Valley for later type stars (mid-K to mid-M; $T_{\mathrm{eff}} < 4700 \unit{K}$) using data from \textit{Kepler} and \textit{K2}, obtaining $r_{\mathrm{transition,CM20}} \propto P^{0.058\pm0.022}$. Their measurement has a power-law slope with the opposite sign to the power-law slope measured by \cite{martinez2019} around Sun-like stars, and is more consistent with models predicting that small planets represent a population of planets that form late in a gas-poor environment \citep{lee2014,lee2016,lopez2018}. \cite{cloutier2020} interpret this that either planet formation is governed by a separate process around M-dwarfs (i.e., gas-poor formation) or that the efficiency of atmospheric post-processing (such as photoevaporation) is weakened for planets orbiting low-mass stars. 

\begin{figure}[t!]
\begin{center}
\includegraphics[width=1.0\columnwidth]{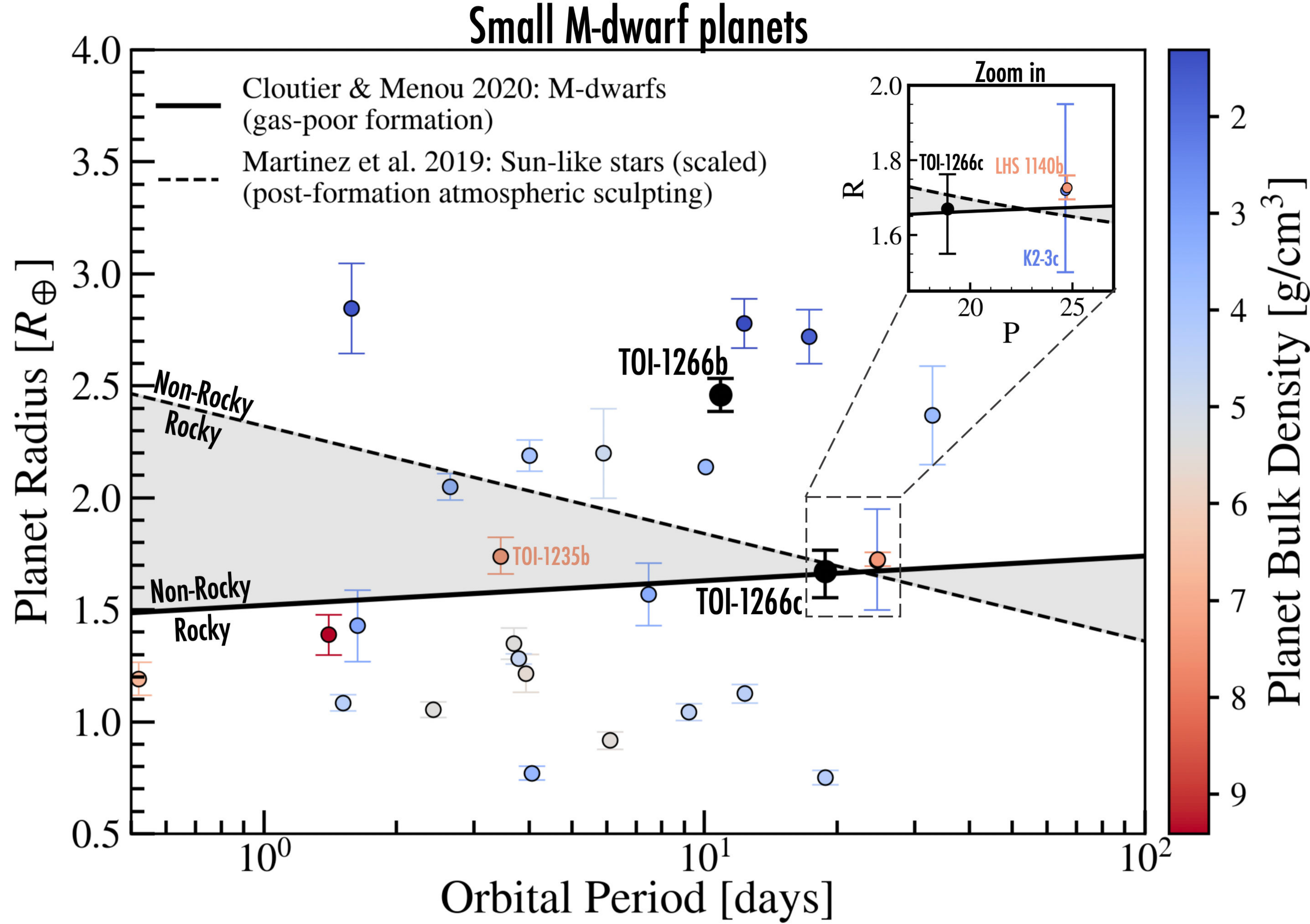}
\end{center}
\caption{Planet radius for small M-dwarf planets ($R<4R_\oplus$) as a function of orbital period. TOI-1266b and c are shown with the black points. Planets with better than 50\% mass constraints have their bulk density highlighted with the color gradient. The solid black line shows the location of the rocky-to-gaseous transition radius $r_\mathrm{transition}$ for planet host stars with $T_{\mathrm{eff}} < 4700 \unit{K}$ as measured by \cite{cloutier2020}, consistent with the predictions of gas-poor formation. The dashed line shows $r_\mathrm{transition}$ as a function of orbital period as measured by \cite{martinez2019} around solar-type stars (scaled to the low-mass regime), consistent with the predictions of photoevaporation or core-powered mass loss models. TOI-1266c lies in the transition region as predicted by \cite{cloutier2020} and \cite{martinez2019}, and could thus have either a predominantly rocky or non-rocky composition. The inset highlights the position of TOI-1266c and two other M-dwarf planets also residing in the Radius Valley: K2-3c and LHS 1140b, which are observed to be gaseous and rocky, respectively. Data obtained from the NASA Exoplanet Archive on May 20, 2020 \citep{akeson2013}.}
\label{fig:rp}
\end{figure}

In Figure \ref{fig:rp}, we show planet radius as a function of orbital period for small ($R<4R_\oplus$) M-dwarf planets with mass measurements better than $50\%$\footnote{We note here that imposing a mass constraint introduces an observational bias as lower mass planets are less likely to have good fractional mass precision.}, which we compare to the Radius Valley locations as measured by \cite{martinez2019} around Sun-like stars, and by \cite{cloutier2020} for M-dwarfs. Following, \cite{cloutier2020}, in Figure \ref{fig:rp}, we plot the Radius Valley location of \cite{martinez2019} after scaling to the M-dwarf mass regime. Specifically, we plot the radius valley location in $r$-$P$ space, as given by Equations 10 and 11 in \cite{cloutier2020}, as,
\begin{equation}
r_{\mathrm{transition,M19}} = -0.48 \log_{10}(P) + 2.32,
\label{eq:M19}
\end{equation}
for solar type stars, and,
\begin{equation}
r_{\mathrm{transition,CM20}} = 0.11 \log_{10}(P) + 1.52,
\label{eq:CM19}
\end{equation}
for M-dwarf stars. 

From Figure \ref{fig:rp}, with a period of $P=18.8$ days and radius of $1.67 R_\oplus$, we see that TOI-1266c lands in the transition region as predicted by both \cite{cloutier2020} for late K and M-dwarf systems (Equation \ref{eq:CM19}), and by \cite{martinez2019} for Sun-like stars (Equation \ref{eq:M19}). As such, TOI-1266c could have a rocky composition or a predominantly non-rocky composition (e.g., a water rich world or could have retained a few percent H/He atmosphere). The inset in Figure \ref{fig:rp} further highlights the position of TOI-1266c and two other M-dwarf planets also residing in the transition region: K2-3c and LHS 1140b, which interestingly show different bulk compositions. K2-3c has a radius of $1.72 \pm 0.22 R_\oplus$ \citep{crossfield2015k2_3c}, and a mass of $2.1 \pm 1.0M_\oplus$ \citep{kosiarek2019}, and thus has a bulk density of $\rho \sim 3 \unit{g/cm^3}$, suggestive of a non-rocky composition. However, LHS 1140 b \citep{dittmann2017,ment2019} has a radius of $1.727 \pm 0.032 R_\oplus$, a mass of $7.0 \pm 0.9 M_\oplus$, and bulk density of $\rho \sim 7.5 \unit{g/cm^3}$, consistent with a rocky composition.

From Figure \ref{fig:rp}, we also note that both LHS 1140b, and TOI-1235b---a planet recently discovered and characterized by \cite{cloutier2020toi1235} and \cite{bluhm2020}---both have densities consistent with rocky compositions, but both reside above the line measured by \cite{cloutier2020}, where we would have predicted them to have a non-rocky composition. This could suggest that the transition region could lie slightly higher than measured in \cite{cloutier2020}. Another explanation would be that the efficiency of different processes sculpting planetary compositions varies for planets in the transition region, resulting in a continuum of possible compositions. This would be compatible with the trend noted by \cite{fultonpetigura2018} and \cite{cloutier2020}, that the Radius Valley is not completely void of planets, and gets increasingly filled with decreasing stellar masses. As mentioned by \cite{cloutier2020}, this trend has not been firmly tested yet. A precise mass constraint of TOI-1266c, along with other planets residing in the Radius Valley, can directly help place further constraints on this trend.

\subsection{TOI-1266 c: A Potential Super-Venus?}
With a radius of $R=1.67 R_\oplus$ and an incident stellar flux 2.4 times that of Earth, if TOI-1266c has a rocky composition, it could potentially be a `Super-Venus' \citep{kane2014}. Venus itself receives 1.91 times more flux and is $95\%$ the size of the Earth. \cite{kane2014} define Venus analogs as predominantly rocky planets residing in the `Venus-Zone', where planets receive insolation fluxes between $\sim$0.95-25 times that of Earth\footnote{The exact bounding values of the Venus-Zone are dependent on the effective temperature of the host star, see Figure 3 in \cite{kane2014}. We have focused here on the bounding values \cite{kane2014} report for M-dwarf systems.}. Future studies attempting to identify atmospheric abundances of small rocky planets will face the challenge of distinguishing between possible Venus and Earth surface conditions \citep{kane2014}. There is a need to discover more planets that may have evolved into a post-runaway greenhouse state so that we can target their atmospheres for characterization with future facilities such as JWST.

As discussed in Subsection \ref{sec:mr} and in Figure \ref{fig:mr}, it is also possible that TOI-1266c could have retained a H/He atmosphere, and/or have a higher water fraction than Earth. If TOI-1266c is determined to be a water-rich world, it remains to be seen how much of it is retained due to the high luminosity pre-main-sequence evolution of its M-dwarf host star \citep{lugerbarnes2015}, and any historical stellar activity. We note that TOI-1266c lies firmly on the side of the ``cosmic shoreline'' where the gravitational binding of the atmosphere to the planet is high compared to the insolation-driven escape \cite[see Figure 1 in][]{ZC2017}, hinting that TOI-1266c could retain a water atmosphere. Interestingly, TOI-1266c also lies very close to or on top of the ``H$_{2}$O greenhouse runaway'' region in \citet{ZC2017}. Depending upon on the age of the system, stellar UV activity and the initial water content accumulated at the early stages of the system evolution, and considering the insolation on the planet, TOI-1266c could host a hot/moist water-vapor atmosphere. Such an atmosphere has recently been detected around the mini-Neptune K2-18b \citep{paper12019,paper22019}. Detecting further such atmospheres would provide a valuable data point in capturing systems that are undergoing moist or runaway greenhouse climates, and provide clues to atmospheric evolutionary history similar to that of the evolution of our own terrestrial planets in the solar system. It would also have implications on initial volatile compound inventories for models of planet formation. 

We estimated the applicability of performing transmission spectroscopy on both planets using the Transmission Spectroscopy Metric (TSM) as defined in \cite{kempton2018}. We obtain a fairly large spread of possible TSMs of $53_{-19}^{+51}$ for planet b using our current mass constraint. Although the median value of 53 is formally below the TSM>90 prioritization threshold for mini-Neptune planets with radii larger than $1.5R_\oplus$ recommended by \cite{kempton2018}, a further precise mass constraint is needed to discern the exact value of the TSM. For planet c, the TSM will depend strongly on if the planet has retained a H/He atmosphere or if the planet is predominantly rocky with a minimal atmosphere. In their definition of the TSM, \cite{kempton2018} define the transition between predominantly rocky planets and gaseous mini-Neptunes at $1.5 R_\oplus$. If we assume TOI-1266c to be a characteristic mini-Neptune, we obtain a TSM of $30_{-15}^{+19}$; if we assume it to be rocky, we obtain a TSM of $5_{-2}^{+3}$. As such, the favorability of TOI-1266c for atmospheric characterization depends strongly if it is determined to be predominantly rocky or non-rocky. 

\section{Summary}
\label{sec:summary}
We have presented the discovery and validation of two small planets orbiting the nearby M2 dwarf TOI-1266. The inner planet has a radius of $2.5 R_\oplus$ and an orbital period of 10.9 days. The outer planet has a smaller radius of $R=1.67 R_\oplus$ and period of 18.8 days, residing in the Radius Valley---the transition region between rocky and gaseous planets. From the available photometry and RVs, we see no clear evidence of other planets in the system.

We validate the planetary nature of the two planets using high contrast imaging observations from NESSI/WIYN, along with ground-based transit photometry---including precision diffuser-assisted photometry of the outer planet using the Engineered Diffuser on the ARC 3.5m Telescope at Apache Point Observatory. Using precision near-infrared RVs from the Habitable-zone Planet Finder, we obtain an upper mass limit of $15.9 M_\oplus$ and $6.4 M_\oplus$ at 95\% confidence for planets b and c, respectively. Our current mass constraints hint that planet c could have a predominantly non-rocky composition, which could indicate that planet c is either water-rich and/or could have retained an atmosphere despite its small size, although further precise RV observations are needed to more precisely constrain its composition. Given the brightness of the host star, both planets are amenable for a precise mass constraint with current and upcoming RV instruments. A precise mass estimate of planet c will further constrain models explaining the emergence of the Radius Valley, and the processes that sculpt the compositions and atmospheres of small planets receiving insolations similar to Venus.

\acknowledgments
We thank Josh Winn for useful discussions. This work was partially supported by funding from the Center for Exoplanets and Habitable Worlds. The Center for Exoplanets and Habitable Worlds is supported by the Pennsylvania State University, the Eberly College of Science, and the Pennsylvania Space Grant Consortium. This work was supported by NASA Headquarters under the NASA Earth and Space Science Fellowship Program through grants 80NSSC18K1114. We acknowledge support from NSF grants AST-1006676, AST-1126413, AST-1310885, AST-1517592, AST-1310875, AST-1910954, AST-1907622, AST-1909506, the NASA Astrobiology Institute (NAI; NNA09DA76A), and PSARC in our pursuit of precision radial velocities in the NIR. Computations for this research were performed on the Pennsylvania State University’s Institute for Computational \& Data Sciences (ICDS).

These results are based on observations obtained with the Habitable-zone Planet Finder Spectrograph on the Hobby-Eberly Telescope. We thank the Resident astronomers and Telescope Operators at the HET for the skillful execution of our observations of our observations with HPF. The Hobby-Eberly Telescope is a joint project of the University of Texas at Austin, the Pennsylvania State University, Ludwig-Maximilians-Universität München, and Georg-August Universität Gottingen. The HET is named in honor of its principal benefactors, William P. Hobby and Robert E. Eberly. The HET collaboration acknowledges the support and resources from the Texas Advanced Computing Center. 

The WIYN Observatory is a joint facility of the University of Wisconsin-Madison, Indiana University, Yale University, the NSF Optical Infrared Research Lab, the University of Missouri, Purdue University, Penn State University, and the University of California at Irvine. Some of the observations in the paper made use of the NN-EXPLORE Exoplanet and Stellar Speckle Imager (NESSI). NESSI was funded by the NASA Exoplanet Exploration Program and the NASA Ames Research Center. NESSI was built at the Ames Research Center by Steve B. Howell, Nic Scott, Elliott P. Horch, and Emmett Quigley. These results are based on observations obtained with the Apache Point Observatory 3.5-meter telescope which is owned and operated by the Astrophysical Research Consortium. We wish to thank the APO 3.5m telescope operators in their assistance in obtaining these data. Some observations were obtained with the Samuel Oschin 48-inch Telescope at the Palomar Observatory as part of the ZTF project. ZTF is supported by the NSF under Grant No. AST-1440341 and a collaboration including Caltech, IPAC, the Weizmann Institute for Science, the Oskar Klein Center at Stockholm University, the University of Maryland, the University of Washington, Deutsches Elektronen-Synchrotron and Humboldt University, Los Alamos National Laboratories, the TANGO Consortium of Taiwan, the University of Wisconsin at Milwaukee, and Lawrence Berkeley National Laboratories. Operations are conducted by COO, IPAC, and UW.

We acknowledge the use of TESS Alert data, from pipelines at the TESS Science Office and at the TESS Science Processing Operations Center. This research has made use of the Exoplanet Follow-up Observation Program website, which is operated by the California Institute of Technology, under contract with the National Aeronautics and Space Administration under the Exoplanet Exploration Program. This paper includes data collected by the TESS mission, which are publicly available from the Multimission Archive for Space Telescopes (MAST). Support for MAST for non-HST data is provided by the NASA Office of Space Science via grant NNX09AF08G and by other grants and contracts. This research made use of Lightkurve, a Python package for Kepler and TESS data analysis (Lightkurve Collaboration, 2018). This research made use of the NASA Exoplanet Archive, which is operated by the California Institute of Technology, under contract with the National Aeronautics and Space Administration under the Exoplanet Exploration Program. This work has made use of data from the European Space Agency (ESA) mission {\it Gaia} (\url{https://www.cosmos.esa.int/gaia}), processed by the {\it Gaia} Data Processing and Analysis Consortium (DPAC, \url{https://www.cosmos.esa.int/web/gaia/dpac/consortium}). Funding for the DPAC has been provided by national institutions, in particular the institutions participating in the {\it Gaia} Multilateral Agreement.

\facilities{TESS, Gaia, HPF/HET 10m, ARCTIC/ARC 3.5m, NESSI/WIYN 3.5m, Perkin 0.4m, ZTF, ASAS.} 
\software{AstroImageJ \citep{collins2017}, 
\texttt{astroplan} \citep{morris2018},
\texttt{astropy} \citep{astropy2013},
\texttt{astroquery} \citep{astroquery},
\texttt{barycorrpy} \citep{kanodia2018}, 
\texttt{batman} \citep{kreidberg2015},
\texttt{corner.py} \citep{dfm2016}, 
\texttt{celerite} \citep{Foreman-Mackey2017}, 
\texttt{dynesty} \citep{speagle2019}, 
\texttt{EXOFASTv2} \citep{eastman2017},
\texttt{exoplanet} \citep{exoplanet2020}, 
\texttt{forecaster} \citep{chen2017}, 
\texttt{GALPY} \citep{bovy2015}, 
\texttt{GNU Parallel} \citep{Tange2011}, 
\texttt{HxRGproc} \citep{ninan2018},
\texttt{iDiffuse} \citep{stefansson2018b},
\texttt{Jupyter} \citep{jupyter2016},
\texttt{juliet} \citep{Espinoza2018},
\texttt{matplotlib} \citep{hunter2007},
\texttt{MRExo} \citep{kanodia2019},
\texttt{numpy} \citep{vanderwalt2011},
\texttt{pandas} \citep{pandas2010},
\texttt{PyMC3} \citep{exoplanet:pymc3},
\texttt{radvel} \citep{fulton2018},
\texttt{SERVAL} \citep{zechmeister2018},
\texttt{starry} \citep{luger2019},
\texttt{tesscut} \citep{tesscut}.}

\bibliography{references}{}
\bibliographystyle{aasjournal}

\newpage

\appendix

\section{HPF Radial Velocities}

Table \ref{tab:rvs} lists the RVs from HPF and associated activity indicators derived from the HPF spectra used in this work.

\begin{table}[H]
\centering
\caption{HPF RVs used in this work along with the Differential Line Width (dLW), Chromatic Index (CRX), and the line indices for the three Ca II IRT triplet lines (Ca II IRT 1, 2 and 3), along with associated errors.}
\begin{tabular}{l c c c c c c}
\hline\hline
BJD &    RV [$\unit{m\:s^{-1}}$] &   dLW [$\unit{m^{2}\:s^{-2}}$]&   CRX  [$\unit{m\:s^{-1}\:Np^{-1}}$]&  Ca II IRT 1 &  Ca II IRT 2 &  Ca II IRT 3 \\ \hline
 2458854.02373 &    $-8.7 \pm 8.1$ &   $69.5 \pm 18.9$ &   $-205.6 \pm 86.9$ &  $0.550 \pm 0.003$ &  $0.308 \pm 0.002$ &  $0.329 \pm 0.002$ \\
 2458859.02978 &    $-7.9 \pm 9.2$ &   $-7.2 \pm 21.9$ &   $290.1 \pm 148.0$ &  $0.570 \pm 0.003$ &  $0.308 \pm 0.003$ &  $0.327 \pm 0.003$ \\
 2458861.02062 &    $1.5 \pm 10.8$ &   $-4.1 \pm 25.4$ &  $-178.4 \pm 157.4$ &  $0.551 \pm 0.004$ &  $0.298 \pm 0.004$ &  $0.320 \pm 0.003$ \\
 2458868.00484 &   $-14.9 \pm 5.9$ &   $21.0 \pm 13.8$ &     $67.8 \pm 77.1$ &  $0.570 \pm 0.003$ &  $0.307 \pm 0.002$ &  $0.328 \pm 0.002$ \\
 2458870.04313 &    $1.1 \pm 11.8$ &   $52.6 \pm 27.8$ &   $211.7 \pm 175.4$ &  $0.549 \pm 0.004$ &  $0.306 \pm 0.004$ &  $0.321 \pm 0.004$ \\
 2458882.99010 &   $17.6 \pm 12.5$ &  $-29.0 \pm 29.3$ &    $21.7 \pm 212.2$ &  $0.562 \pm 0.004$ &  $0.302 \pm 0.004$ &  $0.329 \pm 0.004$ \\
 2458893.92190 &    $-8.4 \pm 7.4$ &   $26.1 \pm 17.6$ &     $48.1 \pm 77.6$ &  $0.553 \pm 0.002$ &  $0.307 \pm 0.002$ &  $0.336 \pm 0.002$ \\
 2458894.92423 &     $3.8 \pm 7.4$ &   $19.4 \pm 17.6$ &   $196.0 \pm 121.2$ &  $0.562 \pm 0.002$ &  $0.303 \pm 0.002$ &  $0.330 \pm 0.002$ \\
 2458939.79500 &    $12.2 \pm 7.1$ &   $30.4 \pm 16.9$ &    $-93.4 \pm 93.8$ &  $0.557 \pm 0.002$ &  $0.303 \pm 0.002$ &  $0.327 \pm 0.002$ \\
 2458940.82880 &  $-14.2 \pm 10.6$ &   $18.7 \pm 25.2$ &   $-227.9 \pm 66.2$ &  $0.557 \pm 0.003$ &  $0.303 \pm 0.003$ &  $0.339 \pm 0.003$ \\
 2458941.80403 &     $5.9 \pm 6.3$ &   $23.4 \pm 15.1$ &     $24.4 \pm 94.6$ &  $0.557 \pm 0.002$ &  $0.300 \pm 0.002$ &  $0.331 \pm 0.002$ \\
 2458967.72965 &    $-4.2 \pm 7.1$ &  $-13.6 \pm 17.0$ &  $-217.5 \pm 107.5$ &  $0.556 \pm 0.002$ &  $0.297 \pm 0.002$ &  $0.327 \pm 0.002$ \\
 2458969.73789 &    $-5.0 \pm 6.9$ &   $21.7 \pm 16.7$ &    $227.4 \pm 62.2$ &  $0.554 \pm 0.002$ &  $0.303 \pm 0.002$ &  $0.319 \pm 0.002$ \\
 2458971.73516 &     $9.3 \pm 7.7$ &   $46.1 \pm 18.3$ &   $-58.6 \pm 118.2$ &  $0.557 \pm 0.002$ &  $0.298 \pm 0.002$ &  $0.317 \pm 0.002$ \\
 2458974.75218 &    $-5.0 \pm 6.9$ &   $36.8 \pm 16.5$ &    $45.2 \pm 111.3$ &  $0.561 \pm 0.002$ &  $0.300 \pm 0.002$ &  $0.322 \pm 0.002$ \\
 2458975.72176 &    $15.0 \pm 7.9$ &   $17.5 \pm 18.9$ &    $54.5 \pm 113.7$ &  $0.557 \pm 0.002$ &  $0.296 \pm 0.002$ &  $0.326 \pm 0.002$ \\
 2458976.71658 &     $2.2 \pm 7.4$ &   $39.3 \pm 17.8$ &   $-40.4 \pm 108.5$ &  $0.559 \pm 0.002$ &  $0.305 \pm 0.002$ &  $0.336 \pm 0.002$ \\
 2458978.70378 &     $9.0 \pm 6.4$ &   $42.5 \pm 15.3$ &     $12.6 \pm 89.4$ &  $0.558 \pm 0.002$ &  $0.300 \pm 0.002$ &  $0.337 \pm 0.002$ \\
 2458979.69504 &    $-5.9 \pm 6.4$ &  $-12.0 \pm 15.5$ &   $-54.6 \pm 118.5$ &  $0.561 \pm 0.002$ &  $0.299 \pm 0.002$ &  $0.333 \pm 0.002$ \\
 2459001.64772 &    $-0.4 \pm 6.7$ &    $2.6 \pm 16.2$ &    $-85.6 \pm 45.3$ &  $0.553 \pm 0.002$ &  $0.294 \pm 0.002$ &  $0.316 \pm 0.002$ \\
 2459002.67916 &   $-12.0 \pm 7.5$ &    $3.6 \pm 18.0$ &    $90.5 \pm 106.1$ &  $0.549 \pm 0.002$ &  $0.292 \pm 0.002$ &  $0.312 \pm 0.002$ \\
 2459003.64601 &     $8.1 \pm 6.7$ &   $-5.4 \pm 16.1$ &     $65.3 \pm 73.3$ &  $0.555 \pm 0.002$ &  $0.293 \pm 0.002$ &  $0.313 \pm 0.002$ \\
\hline
\end{tabular}
\label{tab:rvs}
\end{table}

\end{document}